\documentclass[aps,twocolumn,prc,floatfix,showpacs,preprintnumbers,amsmath,amssymb,nofootinbib,groupedaddress]{revtex4-1}
\usepackage[normalem]{ulem}
\usepackage{wasysym}
\usepackage{color}
\usepackage{graphicx}
\usepackage{dcolumn}    % Align table columns on decimal point
\usepackage{bigstrut}
\begin{document}
% >>>>>>>>>>>>>>>>>>>>>>>>>>>>>>>>>>>>>>>>>>>>>>>>>>>>>>>>>>>>>>>>>>>>
% TITLE AND AUTHORS.
%
\title{The neutron skin thickness from the measured\\ electric dipole polarizability in $^{68}$Ni, $^{120}$Sn, and $^{208}$Pb}

\author{X. Roca-Maza$^1$, X. Vi\~nas$^2$, M. Centelles$^2$, B. K. Agrawal$^3$, G.  Col\`o$^1$, N. Paar$^{4,5}$, J. Piekarewicz$^6$, D. Vretenar$^5$}

\email{xavier.roca.maza@mi.infn.it}

\affiliation{$^1$Dipartimento di Fisica, Universit\`a degli Studi di Milano and INFN,  Sezione di Milano, 20133 Milano, Italy.\\
         $^2$Departament d'Estructura i Constituents de la Mat\`eria and Institut de Ci\`encies del Cosmos (ICCUB), Facultat de F\'{\i}sica, Universitat de Barcelona, Diagonal 645, 08028 Barcelona, Spain.\\
         $^3$Saha Institute of Nuclear Physics, 1/AF Bidhannagar, Kolkata 700064, India.\\
         $^4$Department of Physics, University of Basel, Klingelbergstrasse 82, 4056 Basel, Switzerland.\\
         $^5$Department of Physics, Faculty of Science, University of Zagreb, Zagreb, Croatia.\\
         $^6$Department of Physics, Florida State University, Tallahassee, FL 32306, USA}

\date{\today} 

%=============================================================
% ABSTRACT
%=============================================================
\begin{abstract}
The information on the  symmetry energy and its density dependence is deduced by comparing the available data on the electric dipole polarizability $\alpha_D$ of ${}^{68}$Ni, ${}^{120}$Sn, and ${}^{208}$Pb with the predictions of the Random Phase Approximation, using a representative set of nuclear energy density functionals. The  calculated values of $\alpha_D$ are used  to validate different correlations involving $\alpha_D$, the symmetry energy at the saturation  density $J$, the corresponding slope parameter $L$ and the neutron skin thickness $\Delta r_{\!np}$, as suggested by the Droplet Model. A subset of models that reproduce simultaneously the measured polarizabilities in ${}^{68}$Ni, ${}^{120}$Sn, and ${}^{208}$Pb are employed to predict the values of the symmetry energy parameters at saturation density and $\Delta r_{\!np}$.  The resulting intervals are: $J\!=\!30 \text{-}35$ MeV, $L\!=\!20 \text{-} 66$ MeV; and the values for $\Delta r_{\!np}$ in ${}^{68}$Ni, ${}^{120}$Sn, and ${}^{208}$Pb are in the ranges: 0.15\text{-}0.19 fm, 0.12\text{-}0.16 fm, and 0.13\text{-}0.19 fm, respectively. The strong correlation between the electric dipole polarizabilities of two nuclei is instrumental to predict the values of electric dipole polarizabilities in other nuclei. 
\end{abstract}


%=============================================================
% PACS DESCRIPTION
%=============================================================
%24.30.Cz Giant resonances 
%21.60.Jz Nuclear Density Functional Theory and extensions (includes Hartree-Fock and random-phase approximations) 
%21.65.Ef Symmetry energy

\maketitle 

\section{Introduction}

The density dependence of the nuclear symmetry energy plays a critical role in nuclear physics and astrophysics and it is extensively investigated from both theoretical and experimental perspectives. Given that the nuclear symmetry energy is not an observable which can be directly measured, many experiments that measure closely related observables have been designed to extract information about this fundamental quantity. In particular, the neutron skin thickness and the electric dipole polarizability of nuclei have been identified as strong \emph{isovector indicators}\,\cite{reinhard10}. The main focus of the present work is the electric dipole response. 

Different experimental techniques, such as photo-absorption, Coulomb excitation, and proton scattering at very forward angles (where the Coulomb interaction dominates), have been employed to probe the electric dipole response\,\cite{aumann13,savran2013,bracco15}. These electromagnetic reactions are particularly suited for this purpose because, unlike hadronic experiments, they are not hindered by large and uncontrolled uncertainties. In addition to the identification of the prominent Giant Dipole Resonance (GDR), the electric dipole response of neutron-rich nuclei displays a smaller concentration of strength at lower energies, that is commonly referred to as the Pygmy Dipole Strength (PDS)\,\cite{Paar2007}. Data on the PDS have been used in the past to constrain the symmetry energy and to obtain information on the neutron skin thickness of neutron-rich nuclei\,\cite{Piekarewicz:2006ip,*piekarewicz11,klimkiewicz07,carbone10,roca-maza12a,baran15,papakonstantinou15}. In one of the earliest applications of uncertainty quantification to the domain of energy density functionals (EDFs), Reinhard and Nazarewicz carried out a covariance analysis to correlate the neutron skin thickness of ${}^{208}$Pb to the properties of both finite nuclei and infinite nuclear matter\,\cite{reinhard10}. In this way, the electric dipole polarizability, an observable directly related to the \emph{inverse} energy weighted sum rule, was identified as a strong isovector indicator that may be used to constrain the neutron skin thickness of ${}^{208}$Pb.

The electric dipole polarizability of ${}^{208}$Pb was measured at the Research Center for Nuclear Physics (RCNP)\,\cite{tamii11} using polarized proton inelastic scattering at forward angles. By performing a multipole decomposition of the angular distribution and by measuring all polarization transfer observables, it was possible to extract by two independent methods the electric dipole response of ${}^{208}$Pb over a wide range of energies and with high resolution. Taking into account the average of all available data on the electric dipole response in ${}^{208}$Pb up to the pion-production threshold\,\cite{schelhaas1988,veyssiere1970}, a value for the electric dipole polarizability of $\alpha_D({}^{208}\textrm{Pb})=20.1 \pm 0.6$ fm$^3$ was reported\,\cite{tamii11}. Based on the success of the ${}^{208}$Pb experiment, the electric dipole strength of ${}^{120}$Sn was recently measured at RCNP in the interval between 5 and 22 MeV\,\cite{hashimoto15}. Combining this new measurement with existing photo-absorption data up to 135 MeV\,\cite{lepretre1981}, a value of $\alpha_D({}^{120}\textrm{Sn})=8.93 \pm 0.36$ fm$^3$ was obtained\,\cite{hashimoto15}. Finally, turning to exotic nuclei, the electric dipole response of the unstable ${}^{68}$Ni isotope has been recently investigated at GSI using both Coulomb excitation in inverse kinematics and by measuring the invariant mass in the one- and two-neutron decay channels\,\cite{Wieland:2009,rossi13}.  From these measurements, which cover the range between 7.8 and 28.4 MeV, both the giant and pygmy dipole strength have been identified and the dipole polarizability of $\alpha_D({}^{68}\textrm{Ni})=3.40 \pm 0.23$ fm$^3$ has been obtained\,\cite{rossi13}. Note that neither the high nor the low energy tails of the dipole response of ${}^{68}$Ni are experimentally known, so their contribution was not taken into account in the published value of the polarizability.

As already suggested, the electric dipole polarizability may be used to constrain the neutron skin thickness of ${}^{208}$Pb---and ultimately the density dependence of the symmetry energy. In particular, the experimental determination of $\alpha_D({}^{208}\textrm{Pb})$, combined with a covariance analysis performed with an optimized Skyrme functional (``SV-min''), predicted a neutron skin thickness in ${}^{208}$Pb $\Delta r_{\!np}\!=\!0.156^{+0.025}_{-0.021}$\,fm\,\cite{tamii11}. In a subsequent systematic study performed with a large ensemble of both non-relativistic and relativistic EDFs, a neutron skin thickness $\Delta r_{\!np}\!=\!0.168 \pm 0.022$\,fm was estimated \,\cite{piekarewicz12}. By using relations deduced from the Droplet Model (DM), it was noted that the neutron skin thickness is correlated more strongly with the \emph{product} of the electric dipole polarizability and the symmetry energy coefficient at saturation density ($J$), than to the dipole polarizability alone\,\cite{roca-maza13a}. Using this correlation---and some plausible estimates for $J$---a value of $\Delta r_{\!np}\!=\!0.165 \pm 0.026$ fm was obtained\,\cite{roca-maza13a}. Given the strong correlation between the neutron skin thickness of ${}^{208}$Pb and the slope $L$ of the symmetry energy at saturation density, these results favor a relatively soft symmetry energy with $L\!\simeq\!40$\,MeV, even though with fairly large error bars.

Our aim in this paper is to extract possible constraints on the neutron skin thickness and the symmetry energy parameters by means of a combined analysis of all three recent measurements of the electric dipole polarizability in ${}^{68}$Ni, ${}^{120}$Sn, and ${}^{208}$Pb. To start, we perform self-consistent microscopic calculations of the electric dipole polarizability of all three nuclei in Random Phase Approximation (RPA) using a comprehensive set of EDFs. When required, as in the case of ${}^{120}$Sn, pairing correlations in open-shell nuclei are included by using the quasiparticle RPA (QRPA) framework\,\cite{ringschuck}. The calculated values of the electric dipole polarizability are then used to validate the correlation between $\alpha_D J$, the slope of the symmetry energy $L$, and the neutron skin thickness $\Delta r_{\!np}$, as suggested by the DM formula. Having validated these correlations, we then confront our theoretical predictions against the experimental information in order to select a subset of EDFs that reproduce simultaneously the electric dipole polarizability in all three aforementioned nuclei. Finally, using these selected models we obtain estimates for the neutron skin thickness of ${}^{68}$Ni, ${}^{120}$Sn, and ${}^{208}$Pb, as well as constraints on the symmetry energy parameters. We should emphasize that the experimental values of the electric dipole polarizability for ${}^{120}$Sn and ${}^{208}$Pb contain a small, yet non-negligible, amount of contamination at higher energies caused by non-resonant processes; the so-called ``quasi-deuteron'' effect\,\cite{schelhaas1988,lepretre1981}. To be able to directly compare the measured values of $\alpha_D$ against our theoretical predictions, these contributions have to be subtracted from the experimental strength. The contributions from the quasi-deuteron effect were recently determined\,\cite{atsushi}, so the present analysis uses for the first time the corrected values of the measured $\alpha_D$ to determine the corresponding neutron skin thickness of both ${}^{120}$Sn and ${}^{208}$Pb.

The paper is organized as follows. In Sec.\,\ref{theory} we present a short review of the RPA formalism used to compute the electric dipole response. A brief discussion of the DM approach to the electric dipole polarizability and the correlations suggested by it are also addressed. Particularly important is the connection between the extracted experimental dipole polarizability (minus the quasi-deuteron contribution) and the corresponding theoretical results. Next, we discuss in Sec.\,\ref{results} predictions for the electric dipole polarizability of $^{68}$Ni, $^{120}$Sn, and $^{208}$Pb, obtained using a large and representative set of EDFs. In turn, values of the neutron skin thickness for these nuclei and the associated symmetry energy parameters are estimated from the subset of EDFs which reproduce the data on $\alpha_D$ for all three nuclei. We then exploit these findings to provide genuine predictions for the electric dipole polarizability of both ${}^{48}$Ca and ${}^{90}$Zr---nuclei planned to be experimentally investigated in the near future. Finally, we offer our conclusions in Sec.\,\ref{conclusions}.

\section{Formalism}  
\label{theory}

\subsection{Theoretical concepts}
\label{theo2}

The theoretical description of dynamical properties of nuclear systems, such as the electric dipole polarizability, is usually based on the linearization of the time-dependent Hartree or Hartree-Fock equations in a fully self-consistent way. This means that the residual interaction used to compute the linear response of the nuclear system to an external probe is consistent with the interaction used to generate the mean-field ground state. This technique is commonly known as the Random Phase Approximation (RPA)\,\cite{ringschuck}, and is considered to represent an approximate realization of the small amplitude limit of time-dependent density functional theory. This formalism has been extended to include pairing correlations in the Quasi-particle Random Phase Approximation (QRPA). In the present work and for non-relativistic models, we employ a HF-BCS based approach with the same zero-range, surface pairing force that was used in Ref.~\cite{fracasso2005}, and that gives a reasonable reproduction of the experimental odd-even mass differences in the Sn isotopic chain. For the relativistic counterpart we use the finite-range Gogny force D1S in the particle-particle channel~\cite{berger91}.

The electric dipole strength $R(\omega;E1)$ is evaluated within the (Q)RPA framework using the dipole operator
\begin{equation}
 \mathcal{D} = \frac{Z}{A} \sum_{n=1}^N r_n Y_{1M}(\hat r_n) - 
 \frac{N}{A} \sum_{p=1}^Z r_p Y_{1M}(\hat r_p)\;,  
 \label{diop}
\end{equation}
where $N$, $Z$, and $A$ are the neutron, proton, and mass number, respectively, $r_{n(p)}$ indicates the radial coordinate for neutrons (protons), and $Y_{1M}(\hat r)$ is the corresponding spherical harmonic. This definition of the dipole operator eliminates contaminations to the physical response due to the spurious excitation of the center of mass. Details about nuclear (Q)RPA calculations can be found in Refs.\,\cite{reinhard10,colo13,piekarewicz11,ddme2,*ddme}.

With the electric dipole strength as a function of the excitation energy $\omega$, the dipole polarizability $\alpha_D$ may be computed as follows:
\begin{equation}
\alpha_D  = \frac{8\pi e^2}{9} \int_{0}^{\infty}\!\omega^{-1} 
R(\omega;E1)\,d\omega = \frac{8\pi e^2}{9} m_{-1}(E1) \;,
\label{alphad}
\end{equation}
where $m_{-1}(E1)$ is the inverse energy weighted sum rule. Note that although the $m_{-1}$ moment may be obtained from (Q)RPA calculations, the so-called dielectric theorem\,\cite{bohigas1979,capelli2009,hinohara15} also allows to extract $m_{-1}(E1)$ from a constrained ground-state calculation:
\begin{equation} 
m_{-1}(E1) = 
\frac{1}{2}\left.\frac{\partial^2 \langle \lambda \vert \mathcal{H} 
\vert \lambda\rangle}{\partial \lambda^2}\right\vert_{\lambda = 0} , 
\label{cons}
\end{equation}
where the Hamiltonian $\mathcal{H}$ that describes the nuclear system is ``constrained'' by the field $\lambda\mathcal{D}$, and $\mathcal{D}$ is the dipole operator defined in Eq.\,(\ref{diop}).

It is often possible to invoke semi-classical approaches to elucidate the information content of certain physical observables. Although simple, semi-classical arguments reveal in a very transparent way the underlying physics connected with a given observable. In the particular case of the electric dipole polarizability, the $m_{-1}$ moment may be obtained from a constrained calculation based on Eq.\,(\ref{cons}) using the droplet model of Myers and Swiatecki\,\cite{myers74}. In this case the semi-classical approximation to the electric dipole polarizability reads \cite{meyer82}:
\begin{equation} 
\alpha_{D}^{\rm DM} = \frac{\pi e^2}{54} \frac{A \langle r^2\rangle}{J} 
\left(1+\frac{5}{3}\,\frac{9J}{4Q}A^{-1/3}\right), \label{dpdm1} 
\end{equation} 
where $\langle r^2\rangle$ is the mean-square radius of the nucleus and $Q$ is the surface stiffness coefficient that measures the resistance of the system to the formation of a neutron skin\,\cite{myers74}. In keeping with the fact that the ratio $J/Q$ and the slope parameter $L$ display a strong correlation\,\cite{warda09}, this semi-classical result clearly indicates that the electric dipole polarizability is related to properties of the nuclear symmetry energy\,\cite{satula06}. 

Given its isovector character, it is also natural to expect that a semi-classical expression exists relating the neutron skin thickness $\Delta r_{\!np}^{\rm DM}$ to bulk nuclear properties, such as the ratio $J/Q$, the density of nuclear matter at saturation $\rho_{0}\!\equiv\!3/(4\pi r_{0}^{3})$, and the relative neutron excess $I\!=\!(N\!-\!Z)/A$\,\cite{myers1980,centelles09,warda09}. As elaborated in detail in Ref.\,\cite{roca-maza13a}, the simplicity of the DM allows one to relate the electric dipole polarizability to the neutron skin thickness in a nearly analytical way. Indeed, neglecting corrections to the neutron skin thickness due to both the Coulomb field and the surface diffuseness, one finds
\begin{equation}
\alpha_D^{\rm DM} \!\approx\! \frac{\pi e^{2}}{54} 
\frac{A \langle r^2\rangle}{J}\!\left[1\!+\!\frac{5}{2} 
%\frac{\Delta r_{\!np}^{\rm  DM}}{(I\!-\!I_C) \langle r^2\rangle^{1/2}} \right] \,,
\frac{\Delta r_{\!np}^{\rm  DM}}{I\langle r^2\rangle^{1/2}} \right] \,,
\label{dpdm2}
\end{equation}
%%
%where $I_C=e^2 Z/(20 J r_0 A^{1/3})$ is a correction due to the Coulomb interaction. 
For a given heavy nucleus such as ${}^{208}$Pb, the various correction terms, as well as $\langle r^2\rangle$, computed using many different successful EDFs have very similar values\,\cite{centelles10}. Therefore, Eq.\,(\ref{dpdm2}) suggests that the product $\alpha_DJ$---rather than $\alpha_D$ alone---is strongly correlated to the neutron skin thickness of the nucleus\,\cite{roca-maza13a}. Although inspired by the droplet model, the strong correlation $\alpha_DJ$-$\Delta r_{\!np}$ in ${}^{208}$Pb was validated in Ref.\,\cite{roca-maza13a} by performing self-consistent mean-field plus RPA calculations for both neutron skin thickness and electric dipole polarizability using a rather large and representative set of non-relativistic and relativistic models. As a consequence of this correlation, a high precision measurement of the electric dipole polarizability of a nucleus provides critical information on its neutron skin thickness---if $J$ was known. Moreover, by invoking the well-known correlation between $\Delta r_{\!np}$ and $L$\,\cite{brown00,furnstahl02,warda09,centelles10}, important constraints on the density dependence of the symmetry energy may also be obtained. Finally, based on the established correlation between the neutron-skin thickness of two neutron-rich nuclei\,\cite{brown00,Todd:2003xs,sil05}, we anticipate that the tight correlation between $\alpha_DJ$ and $\Delta r_{\!np}$ observed in ${}^{208}$Pb will extend to other medium- to heavy-mass nuclei. If so, then an $\alpha_D(A_1)J$-$\alpha_D(A_2)J$ correlation between two nuclei (of mass $A_1$ and $A_2$) is also expected to emerge. This can also be easily seen from Eq.~(\ref{dpdm1}). This kind of correlation between the polarizabilities of two nuclei will become instrumental later as we confront our predictions against the experimental results in ${}^{68}$Ni, ${}^{120}$Sn, and ${}^{208}$Pb. 

We emphasize that, although the macroscopic DM provides insightful guidance into the correlations between the dipole polarizability, the neutron skin thickness, and the density dependence of the symmetry energy, all calculations reported in Sec.\,\ref{results} are {\sl microscopic} in origin.  We have computed the neutron skin thickness $\Delta r_{\!np}$ in the mean-field (Hartree or Hartree-Fock) approximation and the polarizability $\alpha_D$ as the dipole response of the mean-field ground state, consistent with the (Q)RPA approach.

\subsection{Theory vs experiment}
\label{tve}

To compare the data on the electric dipole  polarizability in  ${}^{68}$Ni, ${}^{120}$Sn, and ${}^{208}$Pb with the corresponding theoretical (Q)RPA values on a quantitative level, the following comments are in order. The electric dipole response of ${}^{68}$Ni has been measured in the energy interval between 7.8 MeV and 28.4 MeV and the dipole polarizability $\alpha_D({}^{68}\textrm{Ni})=3.40 \pm 0.23$ fm$^3$ has been reported \cite{rossi13}. However, to compare with RPA calculations, the measured dipole response has to be extrapolated to lower and higher energy regions to cover the full range between zero and some upper limit at which the contribution to the dipole polarizability becomes negligible (this limit is expected to lie much below the pion production threshold). The strength below the neutron threshold, which was not accessible in the experiment  \cite{rossi13},  is estimated from the tail of a Lorentzian-plus-Gaussian fit to the deconvoluted data \cite{dominic}. The Lorentzian extrapolates the giant dipole resonance to low energies and the Gaussian takes care of the PDS contribution to the strength. In this fit the error is chosen in such a way that the expected value of the total polarizability at zero energy, that is zero, lies within the $2\sigma$ range. For the nucleus ${}^{68}$Ni, this correction associated with the low-energy strength has an estimated value of 0.39 $\pm$ 0.20 fm$^3$. It is expected that the uncertainty accounts for possible deviations of the hypothesis assumed in the extrapolation method. The strength above the upper experimental limit of 28.4 MeV \cite{rossi13} is also extrapolated from the same Lorentzian fit of the GDR strength. Such an extrapolation to energies above 30 MeV implies, in general, a rather small contribution of the dipole strength to the total polarizability. In the case of  ${}^{68}$Ni, such contribution amounts only to 0.09 $\pm$ 0.05 fm$^3$. Therefore, the adopted value of the dipole polarizability for  ${}^{68}$Ni, including the corrections from the extrapolated low-energy and high-energy regions, is $\alpha_D({}^{68}\textrm{Ni})=3.88 \pm 0.31$ fm$^3$  \cite{dominic}. 

In the high energy region above 30 MeV the experimental dipole strength may contain a non-negligible amount of contamination coming from non-resonant processes (the so-called quasi-deuteron effect \cite{schelhaas1988,lepretre1981}). These contributions should be removed from the experimental strength for a direct comparison with theoretical (Q)RPA results. In the case of ${}^{68}$Ni, as already explained, this region was not explored so no correction is needed\footnote{The Lorenzian extrapolation of the GDR tail at high energies is free from quasi-deuteron contaminations.}. For ${}^{120}$Sn this contribution has an estimated value of 0.34 $\pm$ 0.08 fm$^3$ \cite{atsushi,lepretre1981}. Thus, for a comparison with QRPA calculations, the value $\alpha_D({}^{120}\textrm{Sn}) = 8.93 \pm 0.36$ fm$^3$ of Ref.~\cite{hashimoto15} is replaced by $\alpha_D({}^{120}\textrm{Sn}) = 8.59 \pm 0.37$ fm$^3$. For ${}^{208}$Pb, the quasi-deuteron excitations are estimated to contribute to the dipole polarizability with $0.51\pm 0.15$ fm$^3$ \cite{schelhaas1988,atsushi}, which should be subtracted from the data $\alpha_D({}^{208}\textrm{Pb}) =20.1 \pm 0.6$ fm$^3$ by Tamii et al.~\cite{tamii11}. With this correction, the data used for comparison with the theoretical predictions is $\alpha_D({}^{208}\textrm{Pb}) = 19.6 \pm 0.6$ fm$^3$. In addition, it has to be noted that quasi-deuteron excitations, if not properly subtracted, also lead to values of the experimental EWSR which are inaccurate, much more than for the dipole polarizability.

It should also be mentioned that the 1$p$-1$h$ (Q)RPA formalism is not supposed to reproduce the experimental spreading width of the GDR. In order to do this, one should consider the coupling of the simple 1$p$-1$h$ states with more complicated multi-particle multi-hole configurations. At present, one of the effective ways to account for most of the experimental spreading widths is to take into account the coupling to the collective low-lying (mainly surface) vibrations or phonons \cite{niu15,*litvinova15,*lyutorovich15}. This approach that extends beyond the mean-field approximation is not expected to affect significantly the integral properties of the calculated strength. One way to understand it is the following. We assume that we can simulate the coupling with complicated configurations by smearing the (Q)RPA peaks using Lorentzian functions, so that the experimental resonance width is reproduced. In the case of only one Lorentzian function having a width $\Gamma$, it can be easily shown that the (Q)RPA electric dipole polarizability is, at most, reduced by  
\begin{equation}
\Delta \alpha_D \sim - \alpha_D \frac{\Gamma^2}{4E_{\rm x}^2},
\label{width}
\end{equation}
where $E_{\rm x}$ is the peak energy. Using this equation with the measured values of $E_{\rm x}$ and $\Gamma$ for each nucleus, we find that the correction to $\alpha_D$ should be smaller than $\approx$ 2\%.

\begin{figure*}[t]                    
\includegraphics[width=0.475\linewidth,clip=true]{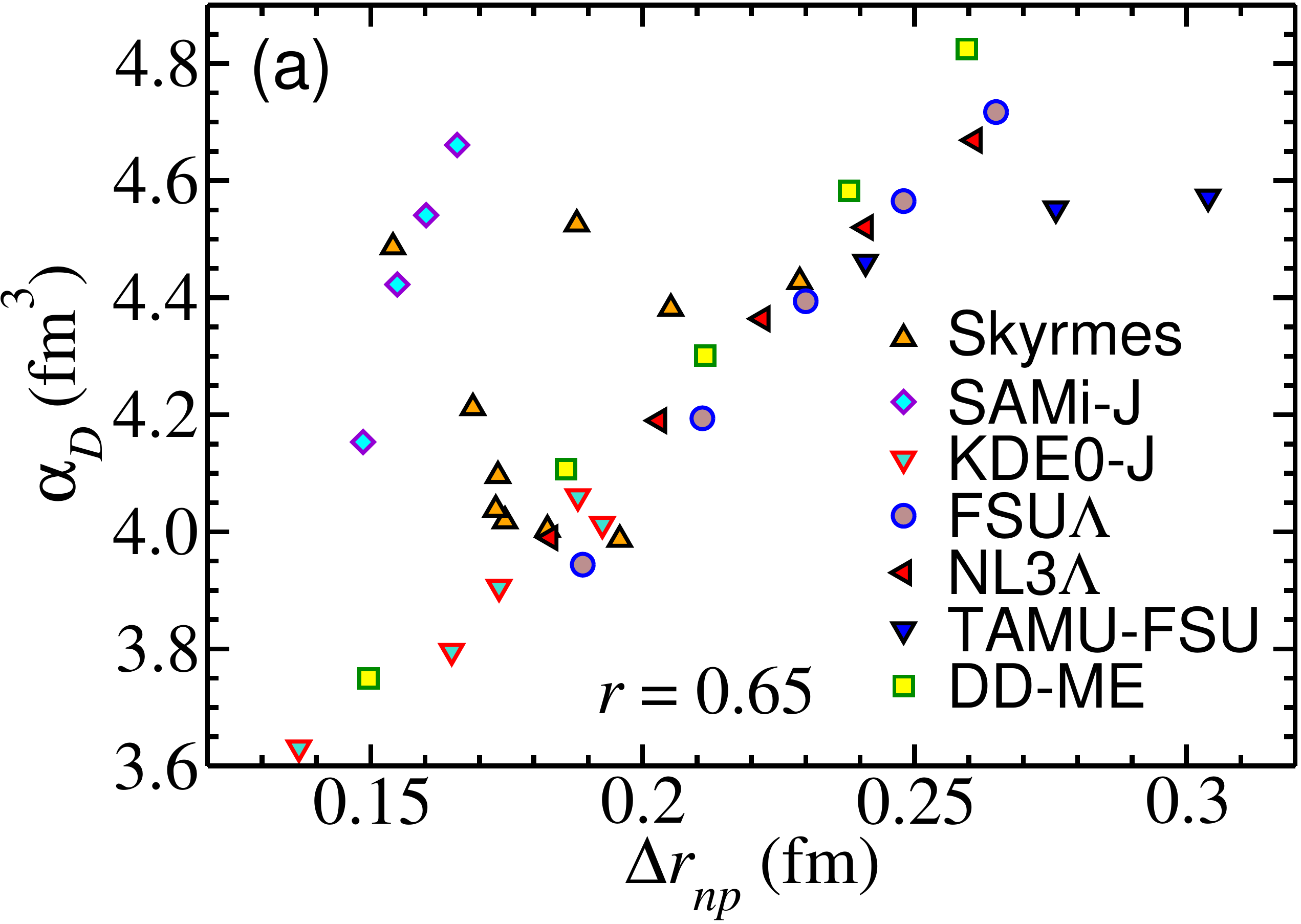}
\hspace{10pt}
\includegraphics[width=0.475\linewidth,clip=true]{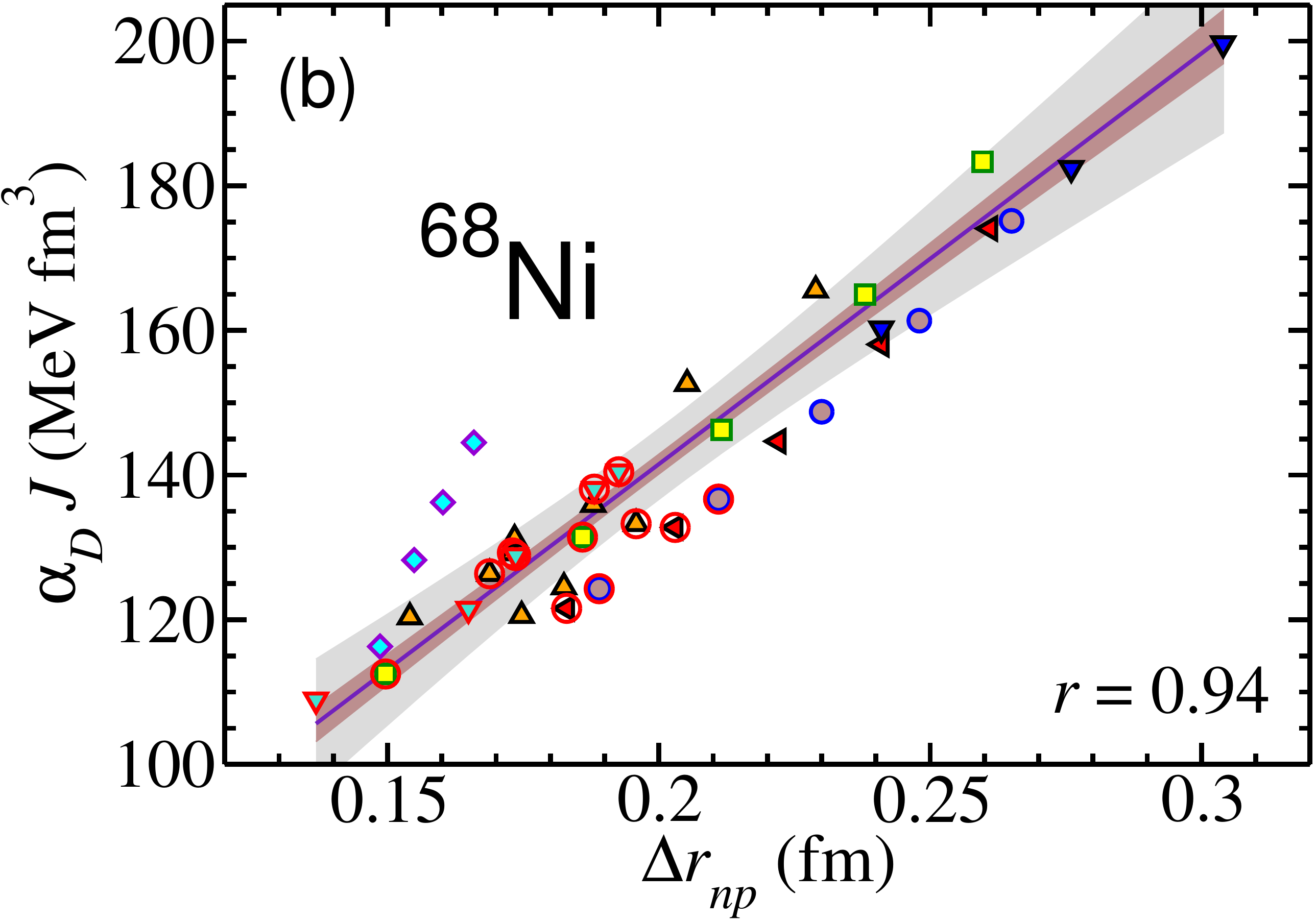}
\caption{(Color online) Plots for the (a) dipole polarizability and (b) product of dipole polarizability times the symmetry energy at saturation $J$ as a function of the neutron-skin thickness for $^{68}$Ni calculated using a large representative set of the EDFs \cite{piekarewicz12,roca-maza12b,*roca-maza13}. Values of $r\!=\!0.65$ and $r\!=\!0.94$ for the respective correlation coefficients are also displayed. The linear fit to the predictions in (b) gives $\alpha_D J = (27\pm15) + (570\pm 33)\Delta r_{\!np}$ and the inner (outer) shadowed regions depict the loci of the 95\% confidence (prediction) bands of regression (see, e.g., Chap. 3 of Ref.~\cite{draper81}). The symbols that are circled in red correspond to those models that are compatible with experiments on the dipole polarizability of both ${}^{68}$Ni and ${}^{208}$Pb.} 
\label{fig1} 
\end{figure*}

\section{Results}
\label{results}

Following the high-resolution measurement carried out at RCNP\,\cite{tamii11}, two systematic studies of the electric dipole polarizability of $^{208}$Pb were performed using a large set of nuclear EDFs\cite{piekarewicz12,roca-maza13a}. Although a robust correlation between the electric dipole polarizability and the neutron skin thickness was found\,\cite{piekarewicz12}, based on the droplet model it was shown that the correlation becomes significantly stronger for the \emph{product} of the electric dipole polarizability and the nuclear symmetry energy coefficient $J$\,\cite{roca-maza13a}. The correlation was indeed found to be very strong, but current uncertainties in the determination of $J$ hinder the determination of the neutron skin thickness of $^{208}$Pb. 

Measurements of the electric dipole polarizability of ${}^{68}$Ni\,\cite{rossi13} and ${}^{120}$Sn\,\cite{hashimoto15} have been recently reported. The aim of this paper is to take full advantage of these experimental developments to constrain both the neutron skin thickness of these nuclei as well as the density dependence of the symmetry energy. Based on this analysis, genuine predictions will be made for the electric dipole polarizability of ${}^{48}$Ca and ${}^{90}$Zr---nuclei that are part of the current experimental campaign at RCNP. 

For our systematic analysis of the electric dipole polarizability we employ a set of non-relativistic Skyrme interactions extensively used in the literature\,\cite{skyrme1,*skyrme2,*lns,*sk255} (these are labeled as ``Skyrmes'' in all the figures). In addition to this set, we employ six different families of systematically varied interactions that are generated by varying their parameters around optimal values, without compromising the quality of the description of well-constrained ground-state properties of finite nuclei, such as masses and charge radii. Two of these families are also based on non-relativistic Skyrme parametrizations; these are labeled in all the figures as SAMi-J\,\cite{roca-maza12b,*roca-maza13} and KDE0-J\,\cite{kde0,kde0j}, respectively. Three of the relativistic families are based on the non-linear Walecka model and are labeled as NL3$\Lambda$, FSU$\Lambda$, and TAMU-FSU\,\cite{Lalazissis:1996rd,nl3fsu1,piekarewicz11,fattoyev13}. Finally, the family labeled as DD-ME corresponds to a relativistic model with density dependent meson-nucleon couplings\,\cite{ddme2,*ddme}.

\subsection{${}^{68}$Ni, ${}^{120}$Sn, and ${}^{208}$Pb}

\begin{figure*}[t]
\includegraphics[width=0.475\linewidth,clip=true]{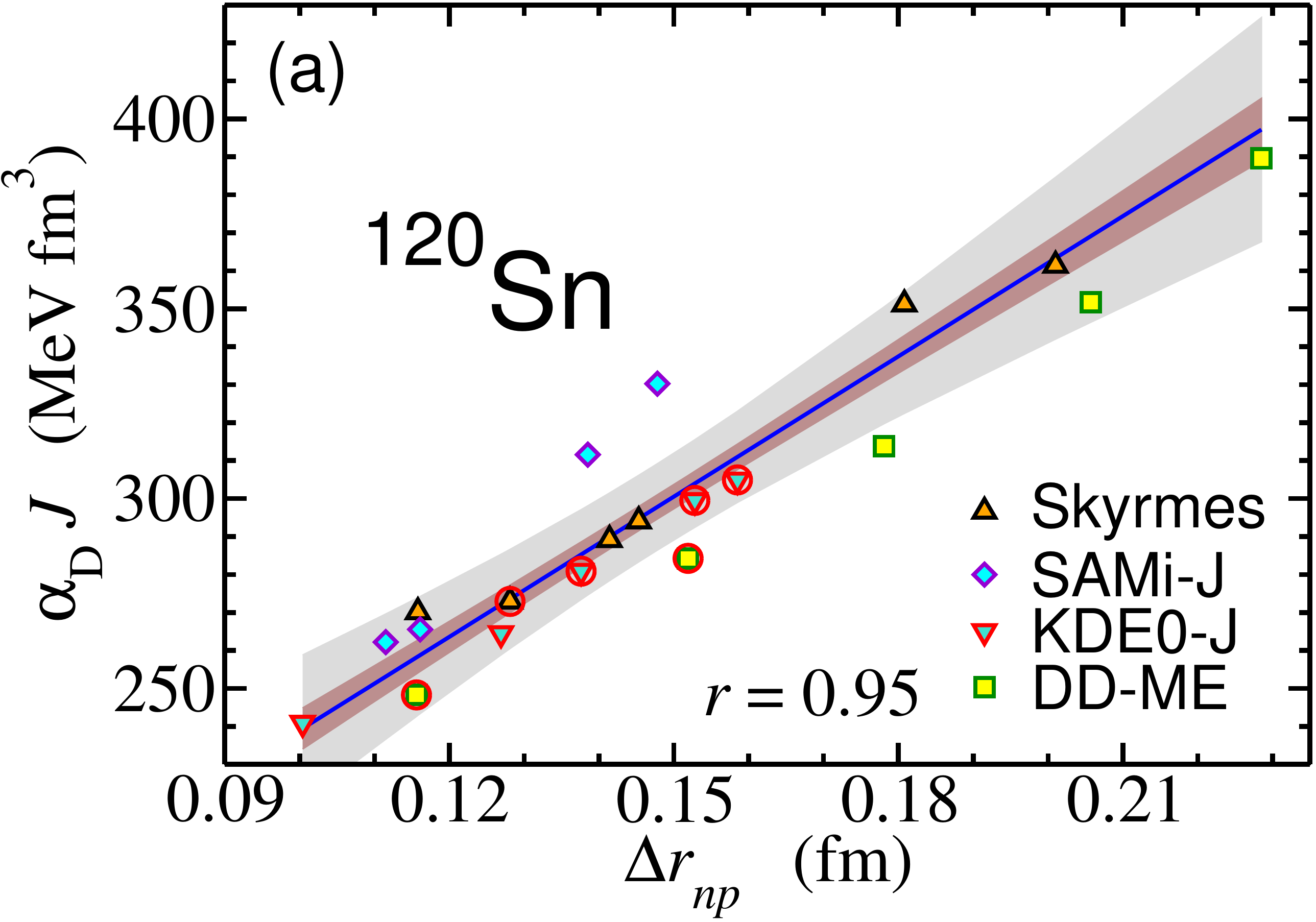}
\hspace{10pt}
\includegraphics[width=0.475\linewidth,clip=true]{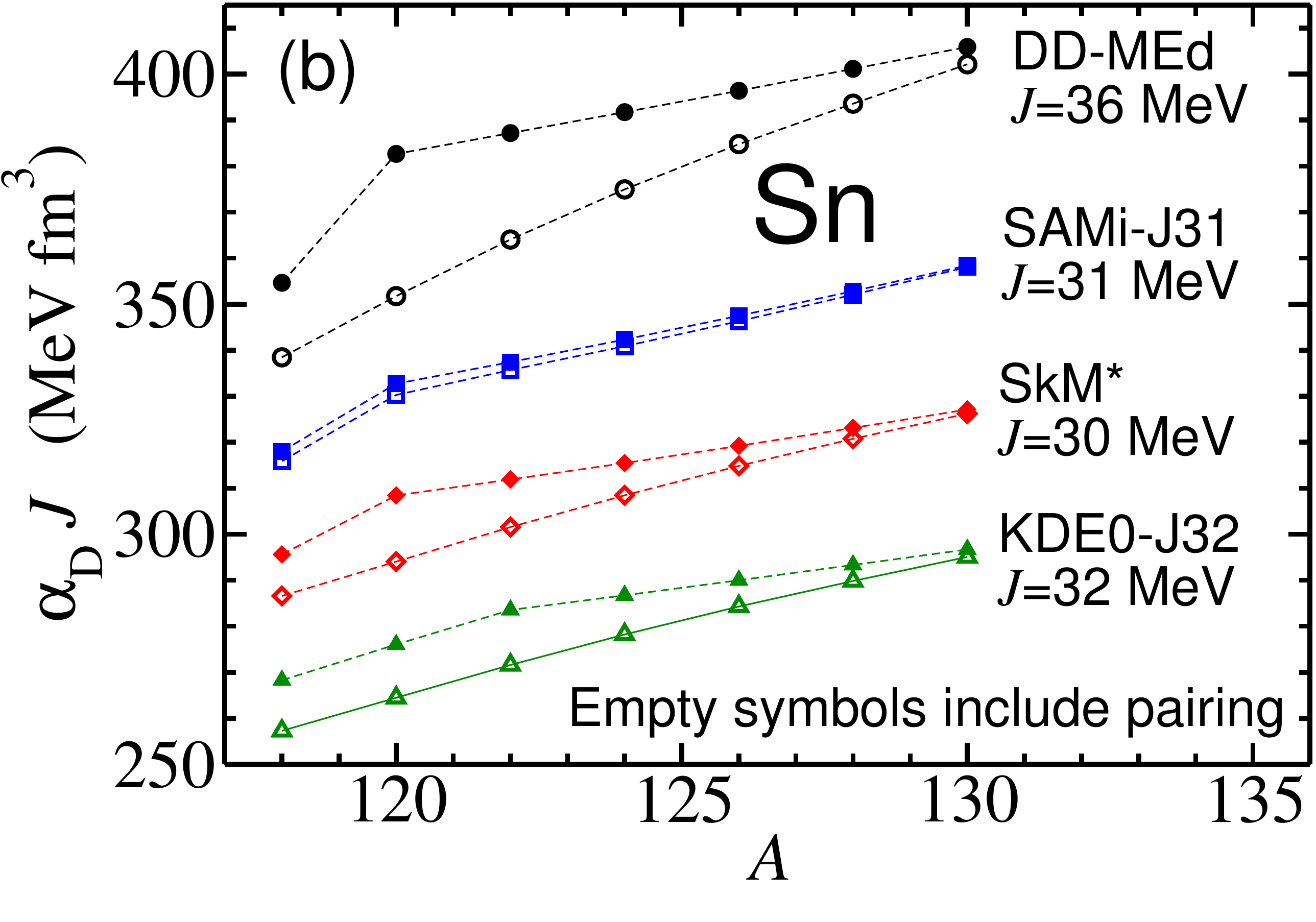}
\caption{(Color on-line) (a) Dipole polarizability times the symmetry energy at saturation $J$ of each EDF against the neutron skin thickness in ${}^{120}$Sn predicted by nuclear EDFs \cite{piekarewicz12,roca-maza12b,*roca-maza13}.  The correlation coefficient is $r=0.95$. The linear fit gives $\alpha_D J = (115\pm 36) + (1234\pm 93)\Delta r_{np}$ and the inner (outer) shadowed regions depict the loci of the 95\% confidence (prediction) bands of the regression (see, e.g., Chap. 3 of Ref.~\cite{draper81}).  The symbols that are circled in red correspond to the models that are compatible with experiments on the dipole polarizability in ${}^{68}$Ni, ${}^{120}$Sn and ${}^{208}$Pb. (b) Dipole polarizability in the even tin isotopes of $A=118-130$ times $J$ as a function of the mass number. The empty (full) symbols correspond to calculations that include (neglect) pairing correlations. In this panel $\alpha_D$ is multiplied by $J$ with the mere purpose to separate the predictions of the different models.} 
 \label{fig2} 
\end{figure*}

We start by displaying in Fig.\,\ref{fig1}a the results for the electric dipole polarizability of ${}^{68}$Ni as a function of neutron skin thickness, as predicted by the large set of EDFs introduced in the previous section. Although a linear correlation may be discerned, a significant amount of scatter among different predictions is clearly observed. Yet, one notes that a linear behavior emerges within each individual family of systematically varied interactions. Overall, the correlation coefficient between $\alpha_D$ and $\Delta r_{np}$ is relatively weak and amounts to only 0.65. However, the correlation coefficient increases considerably---up to 0.96---as soon as the RPA polarizabilities are scaled within each model by the corresponding symmetry energy coefficient $J$; see Fig.\,\ref{fig1}b. This situation is reminiscent of our earlier findings in ${}^{208}$Pb where the correlation coefficient increases from 0.62 to 0.97 upon scaling $\alpha_D$ by $J$\,\cite{roca-maza13a}. We find a similar result for the case of ${}^{120}$Sn. That is, scaling the RPA predictions of $\alpha_D$ by $J$ reduces significantly the model spread. This is observed in Fig.\,\ref{fig2}a where the $\alpha_DJ$-$\Delta r_{np}$ correlation for the nucleus ${}^{120}$Sn is displayed\footnote{Note that we use a reduced set of models, yet representative, as compared to the one displayed in Fig.~\ref{fig1}.}; the inferred correlation coefficient is 0.95. It should be mentioned that we expect that in the open-shell nucleus ${}^{120}$Sn pairing correlations play a non-negligible role as we show below. 

To explore the impact of pairing correlations, we have computed the electric dipole strength of the nucleus ${}^{120}$Sn in the QRPA formalism for a subset of EDFs. In Fig.~\ref{fig2}.b we display the product $\alpha_DJ$, computed with different EDFs, for several tin isotopes as a function of the mass number $A$ without (filled symbols) and with (empty symbols) pairing correlations. The pairing effect on the electric dipole polarizability is more important in mid-shell nuclei and their contribution decreases near magic neutron numbers, as expected. However, the pairing effects can be either very small or large depending on the choice of EDF. In general, pairing reduces electric dipole polarizability in the tin isotopic chain. However, this is not necessarily a systematic effect in all nuclei. In fact, in Ref. \cite{Ebata2014} it has been shown that pairing can at times reduce and at times increase the amount of pygmy dipole strength.

\begin{figure*}[t]
\includegraphics[width=0.475\linewidth,clip=true]{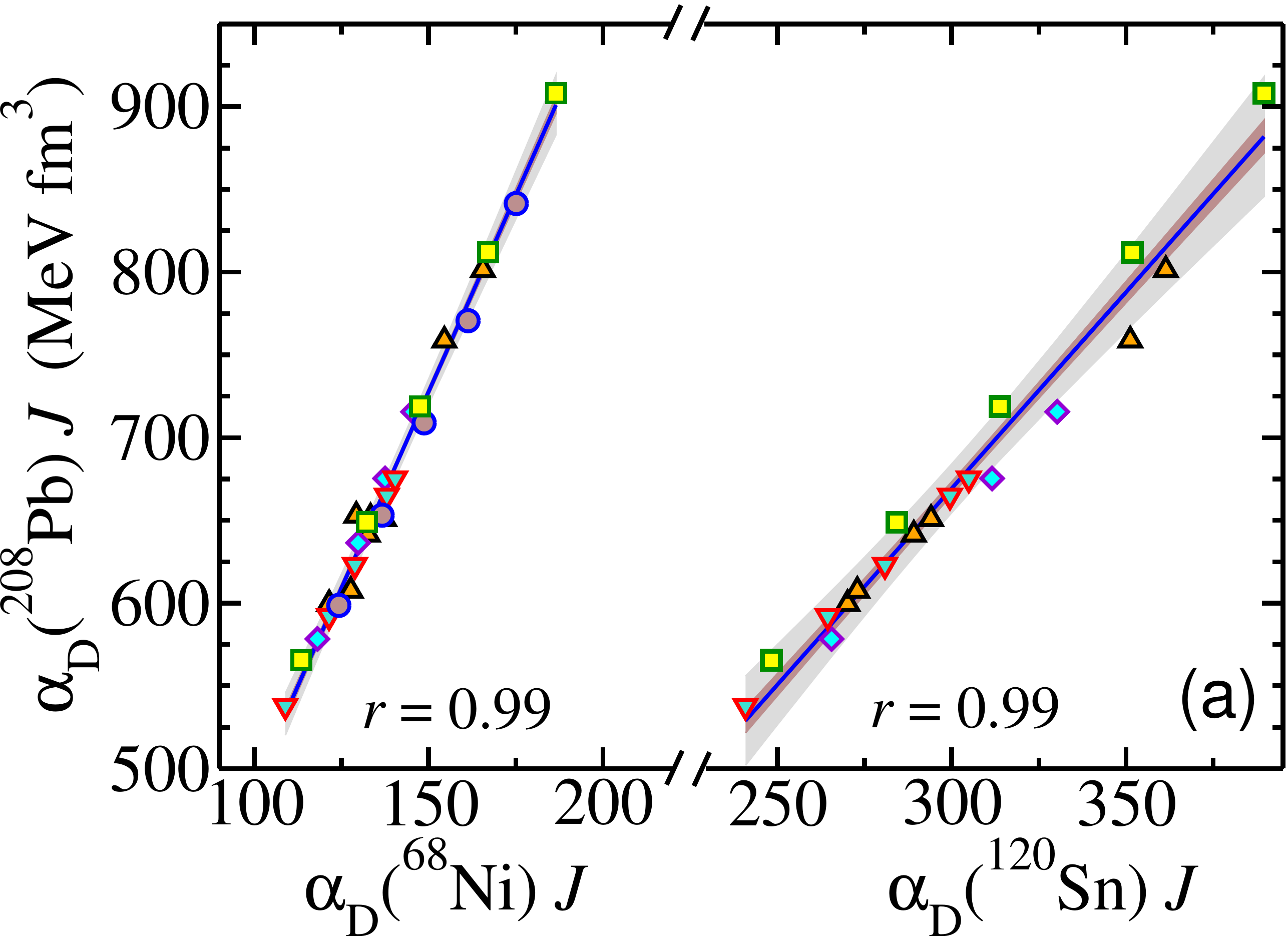}
\hspace{10pt}
\includegraphics[width=0.475\linewidth,clip=true]{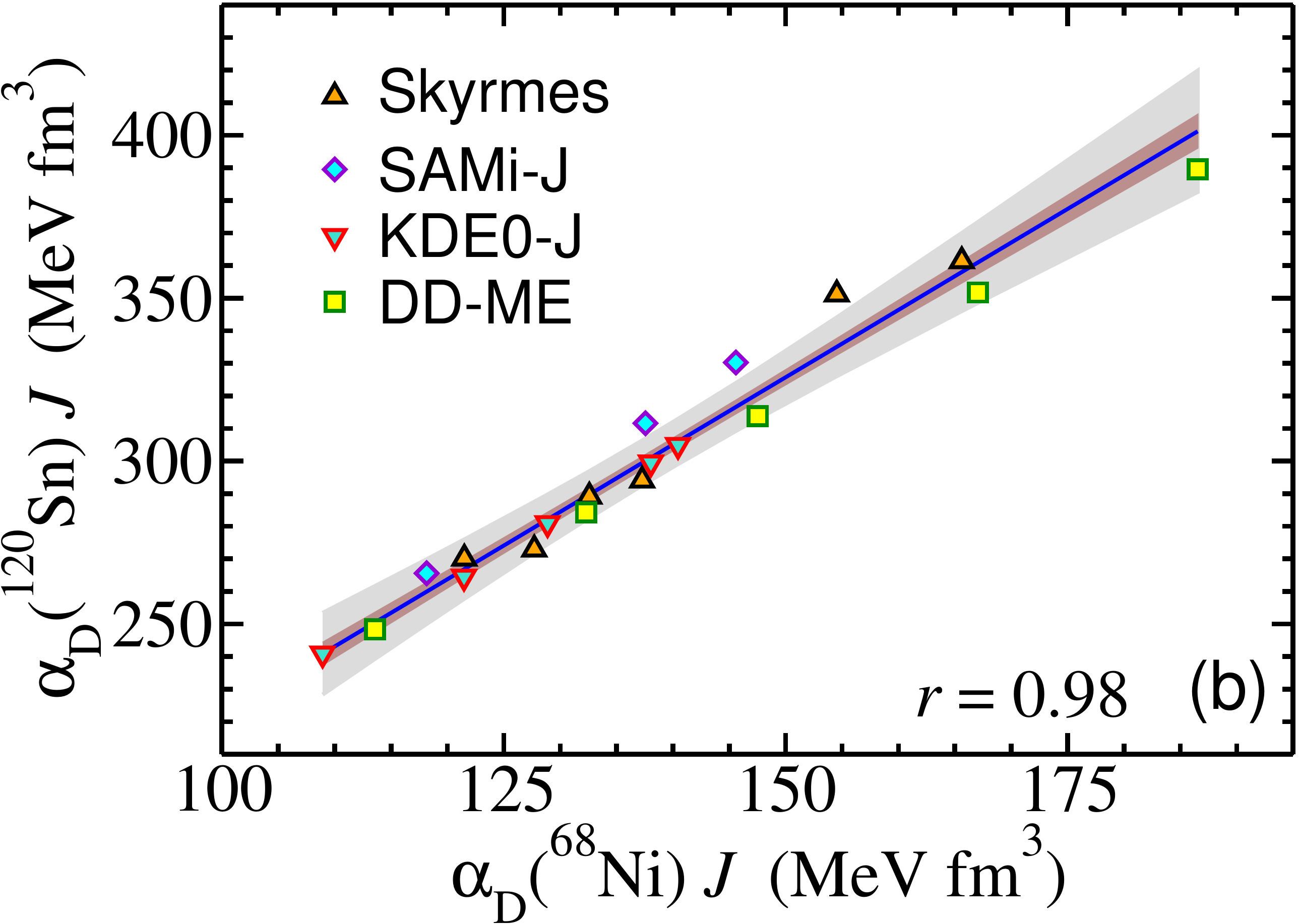}
\caption{(Color on-line) (a) The product $\alpha_DJ$ in  ${}^{208}$Pb against the same product in ${}^{68}$Ni  and ${}^{120}$Sn; in both cases the resulting correlation  coefficients are exceptionally high ($r\!=\!0.99$). The  deduced linear fits give:  $\alpha_D({}^{208}$Pb)$J = (16\pm 2) + (4.7\pm 0.1)\alpha_D({}^{68}$Ni)$J$  and  $\alpha_D({}^{208}$Pb)$J = (-42\pm 4) + (2.4\pm 0.1)\alpha_D({}^{120}$Sn)$J$.  (b) Same as for panel (a) but for the pair ${}^{120}$Sn-${}^{68}$Ni  with a correlation coefficient of $r=0.98$. The linear fit gives  $\alpha_D({}^{120}$Sn)$J = (16\pm 2) + (2.1\pm 0.1)\alpha_D({}^{68}$Ni)$J$.}
\label{fig3} 
\end{figure*}

As already discussed in Sec.\,\ref{theo2}, the correlations implied by the DM formula suggest that the product $\alpha_DJ$ in a given nucleus ($A_{1}$) should be linearly correlated to the same product in another nucleus ($A_2$). To explore the validity of this assertion we display in Fig.\,\ref{fig3}a the linear correlation for the pairs ${}^{208}$Pb-${}^{68}$Ni and ${}^{208}$Pb-${}^{120}$Sn, and for the pair ${}^{120}$Sn-${}^{68}$Ni in Fig.~\ref{fig3}b. The deduced correlation coefficients are exceptionally high---0.99, 0.99, and 0.98, respectively---which confirms the robustness of this correlation.

The use of any correlation involving the product $\alpha_D J$ in a given nucleus to estimate either the neutron skin thickness of the same nucleus (as in Fig.\,\ref{fig1}b) or the dipole polarizability of another nucleus (as in Fig.\,\ref{fig3}) requires knowledge of the symmetry energy coefficient $J$. Indeed, this was the technique employed in Ref.\,\cite{roca-maza13a} to estimate the neutron skin thickness of ${}^{208}$Pb from the measured electric dipole polarizability. To this end, a ``realistic'' value of $J\!=\!31 \pm 2$ MeV was adopted in accordance with two recent analysis\,\cite{tsang12,lattimer2013}; see Ref.\,\cite{roca-maza13a} for further details. However, it should be pointed out that such value of $J$ is deduced from the analysis of different experiments. $J$ is not a physical observable and predictions for the neutron skin thickness and the dipole polarizability---and their associated errors---will be sensitive to the adopted value of $J$. Given that the linear correlations elucidated so far necessarily involve the product $\alpha_DJ$---and that the experimental determination of the dipole polarizability $\alpha_D$ in an increasing number of nuclei is within reach---the need for an accurate determination of $J$ is pressing. Thus, in the following we explore the possibility of constraining $J$, $L$, and $\Delta r_{\!np}$ by comparing the theoretical results to the measured values of the electric dipole polarizability in $^{68}$Ni, $^{120}$Sn, and $^{208}$Pb. Further, these constraints are exploited later so that bona-fide theoretical predictions are provided for the electric dipole polarizability of $^{48}$Ca and $^{90}$Zr, both currently under experimental consideration.

Although scaling $\alpha_D$ by $J$ yields a dramatic improvement in its correlation to $\Delta r_{\!np}$ (see Fig.\,\ref{fig1}), the impact of such scaling in correlating $\alpha_D$ in two different nuclei is far less dramatic. That is, it is possible to estimate the neutron skin thickness of ${}^{68}$Ni, ${}^{120}$Sn, and ${}^{208}$Pb without invoking the empirical value of the symmetry energy $J$. To do so, we identify the subset of accurately calibrated EDFs---out of the large set that we have been employing so far---that simultaneously reproduce the electric dipole polarizability in ${}^{68}$Ni, ${}^{120}$Sn, and ${}^{208}$Pb. These EDFs, which in addition to the electric dipole polarizability reproduce ground-state properties over the entire nuclear chart, provide definite predictions for the neutron skin thickness of the three nuclei, as well as for the two fundamental parameters of the symmetry energy: $J$ and $L$. This approach---now widely adopted by the theoretical community---is reminiscent of a philosophy first proposed by Blaizot and collaborators who advocate a purely microscopic approach for the extraction of nuclear matter parameters ({\sl e.g.,} compression modulus) from the dynamics of giant resonances ({\sl i.e.,} the nuclear breathing mode)\,\cite{Blaizot:1995zz}. While the merit of macroscopic formulas for obtaining qualitative information is unquestionable, the field has attained a level of maturity that demands stricter standards: it is now expected that microscopic models predict simultaneously the strength distribution as well as the properties of nuclear matter.

\begin{figure*}[t!]
\includegraphics[width=0.475\linewidth,clip=true]{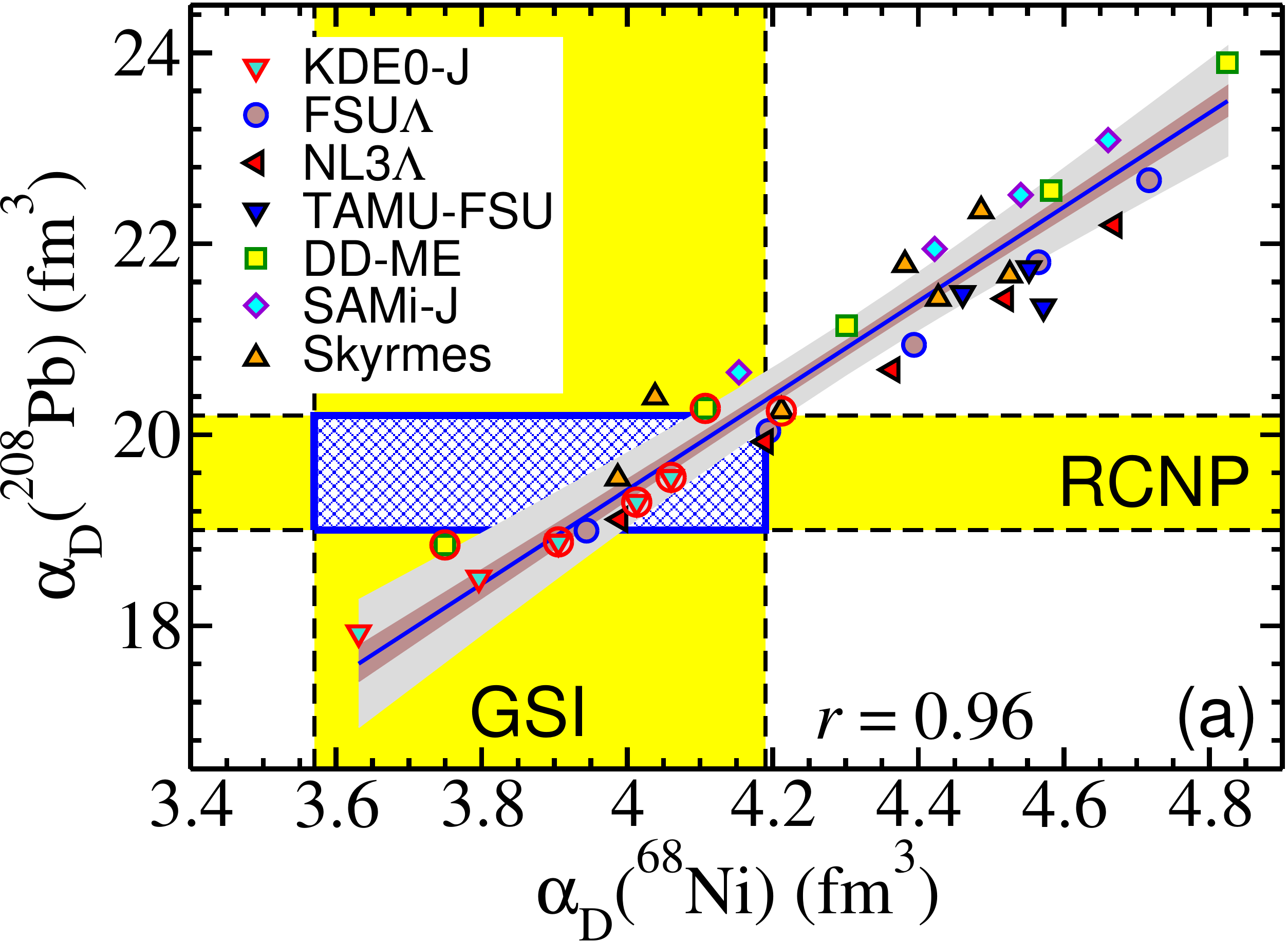}
 \hspace{10pt}
\includegraphics[width=0.475\linewidth,clip=true]{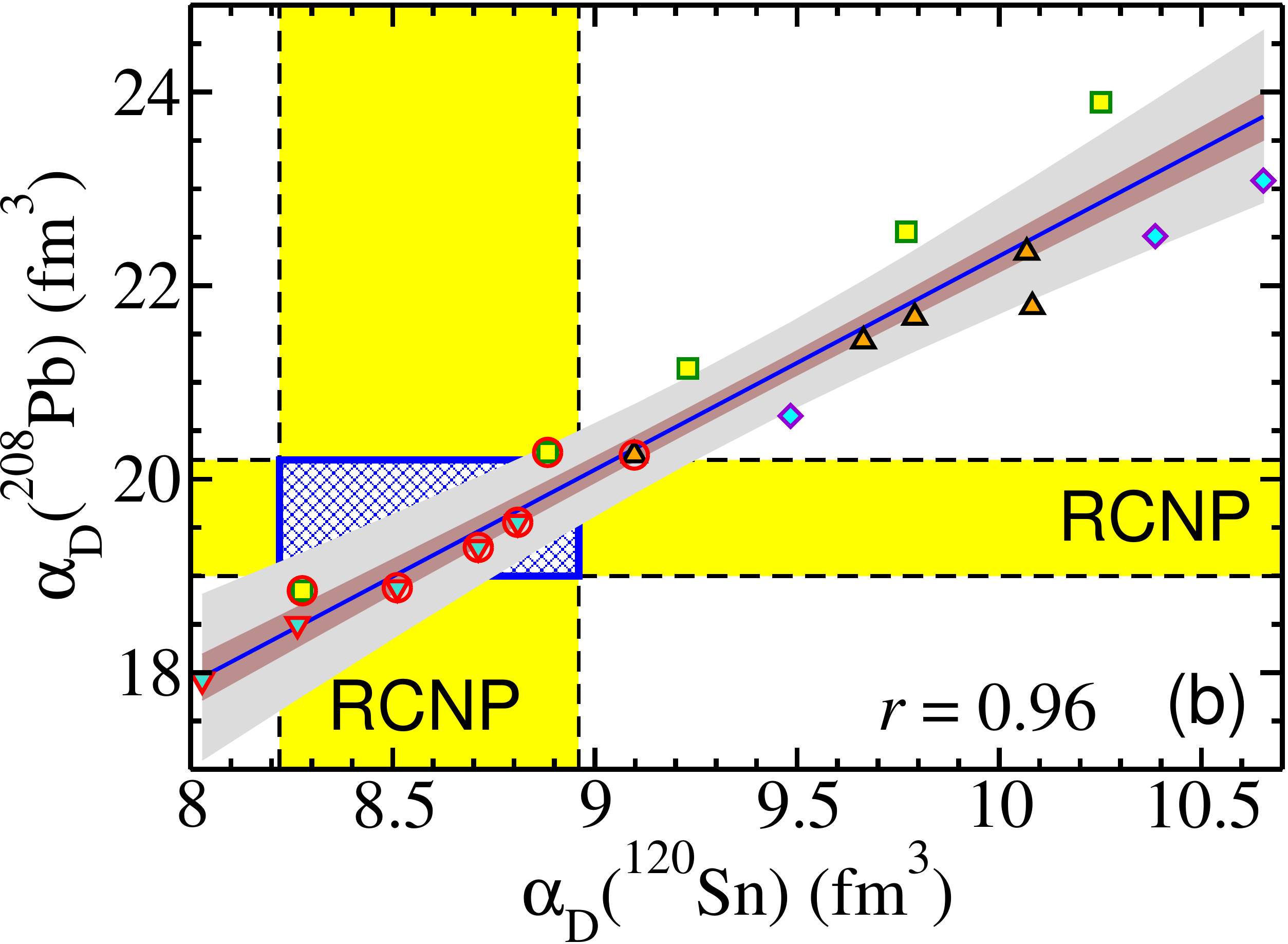}
\caption{Comparison of the theoretical results for the dipole polarizability with the experimental data. (a) ${}^{68}$Ni ($3.88\pm 0.31$ fm$^3$) and ${}^{208}$Pb ($19.6\pm 0.6$ fm$^3$, taking into account the subtraction of the quasi-deuteron excitations $0.51\pm 0.15$ fm$^3$). The linear fit gives $\alpha_D({}^{208}$Pb)$= (-0.5\pm 0.5) + (5.0\pm 0.2)\alpha_D({}^{68}$Ni) and a correlation coefficient $r=0.96$. (b) ${}^{120}$Sn ($8.59\pm 0.37$ fm$^3$, taking into account the subtraction of the quasi-deuteron excitations $0.34\pm 0.08$ fm$^3$) and ${}^{208}$Pb. The linear fit gives $\alpha_D({}^{208}$Pb)$= (0.1\pm 0.5) + (2.2\pm 0.1)\alpha_D({}^{120}$Sn) and a correlation coefficient $r=0.96$. The symbols that are circled in red correspond to the models that are compatible with experiments on the dipole polarizability in ${}^{68}$Ni, ${}^{120}$Sn and ${}^{208}$Pb.}
\label{fig4}
\end{figure*}

We display in Fig.\,\ref{fig4}a and\,\ref{fig4}b the electric dipole polarizability of ${}^{208}$Pb versus those of ${}^{68}$Ni and ${}^{120}$Sn, predicted by the RPA calculation with the set of EDFs used in this work. From the two panels it is seen that $\alpha_D$ in ${}^{208}$Pb remains strongly correlated to $\alpha_D$ in both ${}^{68}$Ni and  ${}^{120}$Sn, although the correlation weakens slightly by removing the scaling with $J$ (see Fig.\,\ref{fig3}). The linear fits obtained from the correlations displayed in Fig.\,\ref{fig4} yield
%%%
\begin{align}
 & \alpha_D({}^{208}\textrm{Pb})=(-0.5\pm 0.5) + 
 (5.0\pm 0.2)\,\alpha_D({}^{68}\textrm{Ni})\,,
 \label{adni} \\
& \alpha_D({}^{208}\textrm{Pb})= (\phantom{-}0.1\pm 0.5)  + 
 (2.2\pm 0.1)\,\alpha_D({}^{120}\textrm{Sn})\,,
\label{adsn}
\end{align}
%%% 
with a correlation coefficient of 0.96 in both cases. Note that to leading order in $A$, Eq.\,(\ref{dpdm1}) largely accounts for the slope between a pair of dipole polarizabilities as predicted by a given interaction---{\sl i.e.,} for fixed values of $J$ and $Q$. That is, $\alpha_D (A_1)\!\sim\!(A_1/A_2)^{5/3} \alpha_D (A_2)$.

Represented by horizontal and vertical yellow bands in the two panels of Fig.\,\ref{fig4} are the experimental values of the electric dipole polarizability, including error bars. It is important to remember that for a quantitative comparison with the theoretical predictions, the experimental values have been corrected as described in Sec.\,\ref{tve}. The theoretical predictions inside the area bounded by the horizontal and vertical bands reproduce the experimental values of both ${}^{208}$Pb and ${}^{68}$Ni, or ${}^{208}$Pb and ${}^{120}$Sn. Red circles emphasize those models that reproduce simultaneously the electric dipole polarizability in all three nuclei. The figure shows that the majority of models that correctly predict the experimental value of $\alpha_D$ in ${}^{208}$Pb are also able to reproduce the data on ${}^{68}$Ni and ${}^{120}$Sn. 

Nevertheless, it should be noted that if the constraint from the measured value of $\alpha_D$ in ${}^{208}$Pb were neglected, {\sl i.e.,} the horizontal yellow band would be omitted from the figure, the experimental values for $\alpha_D$ in ${}^{120}$Sn and, {\sl especially}, in ${}^{68}$Ni, would accommodate more models on the side of softer symmetry energy (smaller $\alpha_D$) and, consequently, on the side of smaller neutron skin thickness. Thus, even {\sl after} applying the corrections described in Sec.~\ref{tve} to the experimental data for $\alpha_D$, which increased the value of $\alpha_D$ in ${}^{68}$Ni and decreased it in ${}^{120}$Sn and ${}^{208}$Pb, it seems that the measured dipole polarizability in the ${}^{68}$Ni nucleus favors a softer nuclear symmetry energy compared to the measurements in ${}^{120}$Sn and ${}^{208}$Pb. 

\begin{table}[tb]
\caption{\label{table1} Various estimates of the neutron skin thickness (in fm) of $^{68}$Ni, $^{120}$Sn, and $^{208}$Pb. (a) Lower and upper values of $\Delta r_{np}$ as predicted by those models that reproduce the experimental values of the electric dipole polarizability of $^{68}$Ni, $^{120}$Sn, and $^{208}$Pb. (b) Mean value and standard deviation of $\Delta r_{np}$ as predicted by the same subset of models in (a). (c) Predictions extracted from the correlation $\alpha_DJ$-$\Delta r_{np}$ using a suitable range for the symmetry energy coefficient $J$ (see text for details).}
\begin{center}
\begin{ruledtabular}
\begin{tabular}{rccc}
Nucleus & $\Delta r_{np}$ (a) & $\Delta r_{np}$ (b) & $\Delta r_{np}$ (c) \\
 \hline\rule{0pt}{2.5ex}
$^{68}$Ni   & 0.15--0.19  & 0.18 $\pm$ 0.01 &  0.16  $\pm$ 0.04\\
$^{120}$Sn  & 0.12--0.16  & 0.14 $\pm$ 0.02 &  0.12  $\pm$ 0.04\\
$^{208}$Pb  & 0.13--0.19  & 0.16 $\pm$ 0.02 &  0.16 $\pm$ 0.03
\end{tabular}
\end{ruledtabular}
\end{center}
\end{table}

A viable option to estimate the neutron skin thickness is to determine an interval using the largest and smallest values predicted by those models that successfully reproduce the experimental dipole polarizabilities in ${}^{68}$Ni, ${}^{120}$Sn, and ${}^{208}$Pb ({\sl cf.} Ref.\,\cite{hashimoto15}). The range of values so obtained is displayed in the first column of Table~\ref{table1}. The second column of the table lists the average values and deviations of the neutron skin thickness predicted by the same subset of selected EDFs. For consistency, we also compare these results with the values extracted using directly the $\alpha_DJ$-$\Delta r_{\!np}$ correlation, as was originally done in Ref.\,\cite{roca-maza13a} for the case of ${}^{208}$Pb. From the correlations displayed on the right panel of Fig.\,\ref{fig1}, on the left panel of Fig.\,\ref{fig2}, and from our previous work on ${}^{208}$Pb one obtains:
%%%
\begin{equation}
 \!\alpha_D J \!=\!
 \begin{cases}
  \phantom{1}(27\pm15) \!+\! (\phantom{1}570\pm33)\Delta r_{\!np}, & \text{for } 
  {}^{68}{\rm Ni}; \\
  (115\pm36) \!+\! (1234\pm93)\Delta r_{np}, & \text{for } {}^{120}{\rm Sn}; \\
  (301\pm32) \!+\! (1922\pm73)\Delta r_{np}, & \text{for } {}^{208}{\rm Pb}. 
 \end{cases}
 \label{fitaD}
\end{equation}
%%%
Given that the extraction of $\Delta r_{\!np}$ from this correlation requires an estimate for the value of $J$, we show here the results obtained by adopting the same choice as in Ref. \cite{roca-maza13a}, namely, $J\!=\!31 \pm 2$\,MeV \cite{tsang12,lattimer2013}. This choice allows one to estimate the neutron skin thickness of ${}^{68}$Ni, ${}^{120}$Sn, and ${}^{208}$Pb using the fits displayed in Eq.\,(\ref{fitaD}). The resulting values for $\Delta r_{\!np}$ in $^{68}$Ni, $^{120}$Sn and $^{208}$Pb are given in the last column of Table\,\ref{table1}. From the results displayed in Table\,\ref{table1}, we notice that the predictions for $\Delta r_{\!np}$ obtained using the subset of EDFs that reproduce the experimental electric dipole polarizabilities of ${}^{68}$Ni, ${}^{120}$Sn, and ${}^{208}$Pb lie within the ranges predicted by the $\alpha_DJ$-$\Delta r_{\!np}$ correlation. This important consistency check suggests that one could in principle use the subset of selected EDFs to predict $\Delta r_{\!np}$ (see column (a) in Table\,\ref{table1}) and then use the tight $\alpha_DJ$-$\Delta r_{\!np}$ correlation to infer a suitable interval of values for $J$ (see below). Note that the neutron skin thickness of ${}^{68}$Ni reported in Ref.\,\cite{rossi13} from the analysis of $\alpha_D$ is $\Delta r_{\!np}\!=\!0.17\pm0.02$\,fm, which is also consistent with the estimates provided in Table\,\ref{table1}. We note that in the analysis that led to this value the authors of Ref.\,\cite{rossi13} compared the experimental dipole polarizability to the RPA calculations within the measured energy interval. A similar analysis was carried out in Ref.\,\cite{hashimoto15} to extract the neutron skin thickness in ${}^{120}$Sn from a measurement of the electric dipole polarizability. The reported value of $\Delta r_{\!np}\!=\!0.148\pm0.034$\,fm in ${}^{120}$Sn again falls within the range predicted in Table\,\ref{table1}, although there is a slight tendency toward the upper limit. In this regard, it is pertinent to point out a difference between the analysis presented here and the one from Ref.\,\cite{hashimoto15}. In Ref.\,\cite{hashimoto15} the contribution from the quasi-deuteron excitations was not subtracted from the data before comparing with QRPA calculations. Finally, for the case of ${}^{208}$Pb the value included in the last column of Table \ref{table1} is consistent the one reported in Ref.\,\cite{roca-maza13a}, {\sl i.e.,} $0.165\pm0.026$\,fm.
   
As noted above, from the present study on the electric dipole polarizability in ${}^{68}$Ni, ${}^{120}$Sn, and ${}^{208}$Pb, we can also obtain information on $J$ and $L$ by choosing the values predicted by the selected set of EDFs that reproduce the experiment in all three nuclei. Following this procedure one obtains the estimates:
%%%
\begin{align}
 & 30 \le J \le 35 \,{\rm MeV}\,, \label{Jrange} \\
 & 20 \le L \le 66\,{\rm MeV}\,.  \label{Lrange}
\end{align}
%%% 
The interval for the symmetry energy is slightly larger than the $J\!=\!31\pm2$\,MeV estimate extracted from a combination of various experiments\,\cite{tsang12,lattimer2013}. The range for the slope of the symmetry energy $L$ predicted by the subset of selected EDFs lies at the lower end of accepted values when compared to other analysis (see, {\sl e.g.,} Refs.\,\cite{dutra12,vinas14,bao13}), yet it is consistent with studies involving giant resonances\,\cite{colo15}. We emphasize that the limits deduced in the present work follow from the analysis of relatively clean electromagnetic experiments. Future electroweak measurements will help narrow these intervals even further.

\begin{figure}[t!]
\includegraphics[width=0.9\linewidth,clip=true]{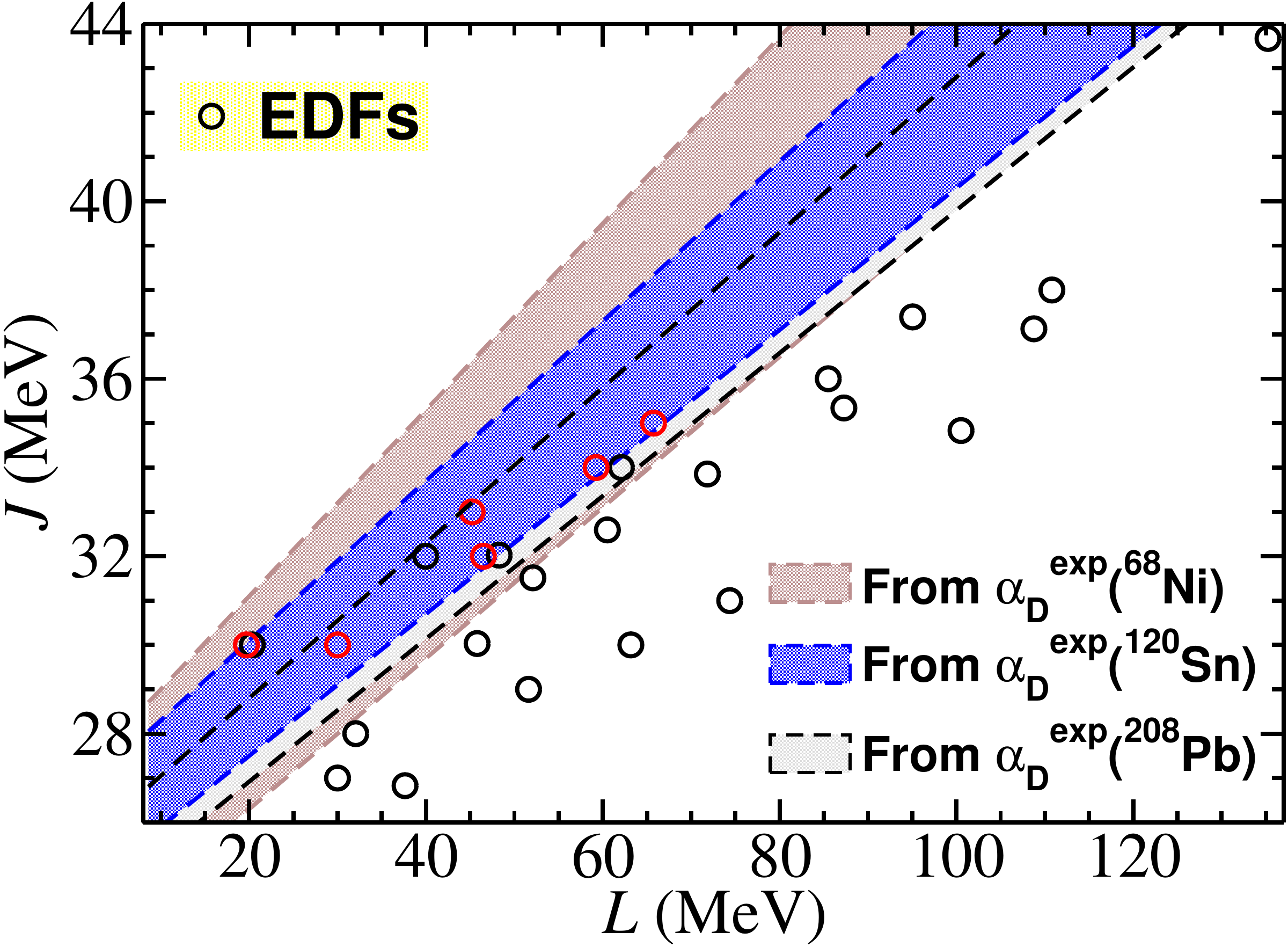}
\caption{$J$ versus $L$ plot showing the constraints obtained in Eqs.~(\ref{jlpb})--(\ref{jlsn}). We also display the predictions of the EDFs employed in this work. We highlight the models that reproduce the experimental $\alpha_D$ in ${}^{68}$Ni, ${}^{120}$Sn, and ${}^{208}$Pb by using red circles.}
\label{fig5}
\end{figure}

Given the strong correlation between the neutron skin thickness of a neutron-rich nucleus and the slope of the symmetry energy $L$\,\cite{centelles09,warda09,roca-maza11}, it is reasonable to expect that the $\alpha_DJ$-$\Delta r_{np}$ correlation will extend to the $\alpha_DJ$-$L$ case, as it has been explicitly shown for ${}^{208}$Pb; see Fig.\,2 of \cite{roca-maza13a} where a correlation $\alpha_D({}^{208}\textrm{Pb}) J = (480\pm 4) + (3.3\pm 0.1)L$ with $r=0.96$ was found. Note that this correlation is also consistent with the DM estimate of $\alpha_{\mathrm{D}}$ (cf. Eq.(11) of \cite{roca-maza13a}). The relation between $J$ and $L$ extracted from this correlation, assuming the experimental value of $\alpha_D({}^{208}\textrm{Pb}) = 19.6 \pm 0.6$ fm$^3$, is
\begin{equation}
J = (24.5\pm 0.8) + (0.168\pm 0.007)L \ .
\label{jlpb}
\end{equation}
The same can be done for ${}^{68}$Ni and ${}^{120}$Sn obtaining in both cases a high correlation for $\alpha_D J - L$ with $r=0.96$. Assuming the experimental values for $\alpha_D$ in these two nuclei, we find 
\begin{eqnarray}
J = (24.9\pm 2.0) + (0.19\pm 0.02)L \label{jlni} \\
J = (25.4\pm 1.1) + (0.17\pm 0.01)L \ , \label{jlsn}
\end{eqnarray}
respectively. We exhibit these constraints as bands in a $J-L$ plot in Fig.~\ref{fig5}. In addition, we display the predictions of the EDFs employed in this work, highlighting those that reproduce the experimental $\alpha_D$ in ${}^{68}$Ni, ${}^{120}$Sn, and ${}^{208}$Pb with red circles\footnote{As an example, the interaction KDE0-J32 with $J=32$ MeV and $L=40$ MeV is compatible with the three bands but not with the experiment on $\alpha_D({}^{208}\textrm{Pb})$. Other interactions depicted in black and compatible with the bands were not tested for the case of ${}^{120}$Sn.}. Our analysis together with the experimental data on the polarizabilities predict three compatible bands with very similar slopes. On the one side, the point of interception with the vertical axis is essentially the same within the error bars (average value of $\approx 24.9$ MeV). This is because it represents the symmetry energy at some average subsaturation density $\langle\rho\rangle$ that has been probed in $\alpha_D$ experiments \cite{tamii11,rossi13,hashimoto15}. To qualitatively understand this, we expand the symmetry energy $S(\rho)$ around the nuclear saturation density $\rho_0$ as $S(\rho) = J - L\epsilon + \mathcal{O}[\epsilon^2]$, where $\epsilon\equiv(\rho_0-\rho)\big/3\rho_0$. Comparing this expansion with Eqs.~(\ref{jlpb})--(\ref{jlsn})---that have the form $J = a + b L$, one can immediately recognize that the ``$a$'' found in the analysis is approximately equal to $S(\langle\rho\rangle)$ and that ``$b$'' allows to roughly estimate the value of $\langle\rho\rangle$. Of course, this interpretation is only valid for small values of ``$b$''. On the other side, the slope of such bands is clearly different from the one depicted by the EDF models. This feature may point towards a possible deficiency in current EDFs: data on $\alpha_D$ impose that a model with a large value of $J$ will need to predict a smaller value of $L$ when compared to the current trend in EDFs.

\begin{figure*}[t!]
\includegraphics[width=0.45\linewidth,clip=true]{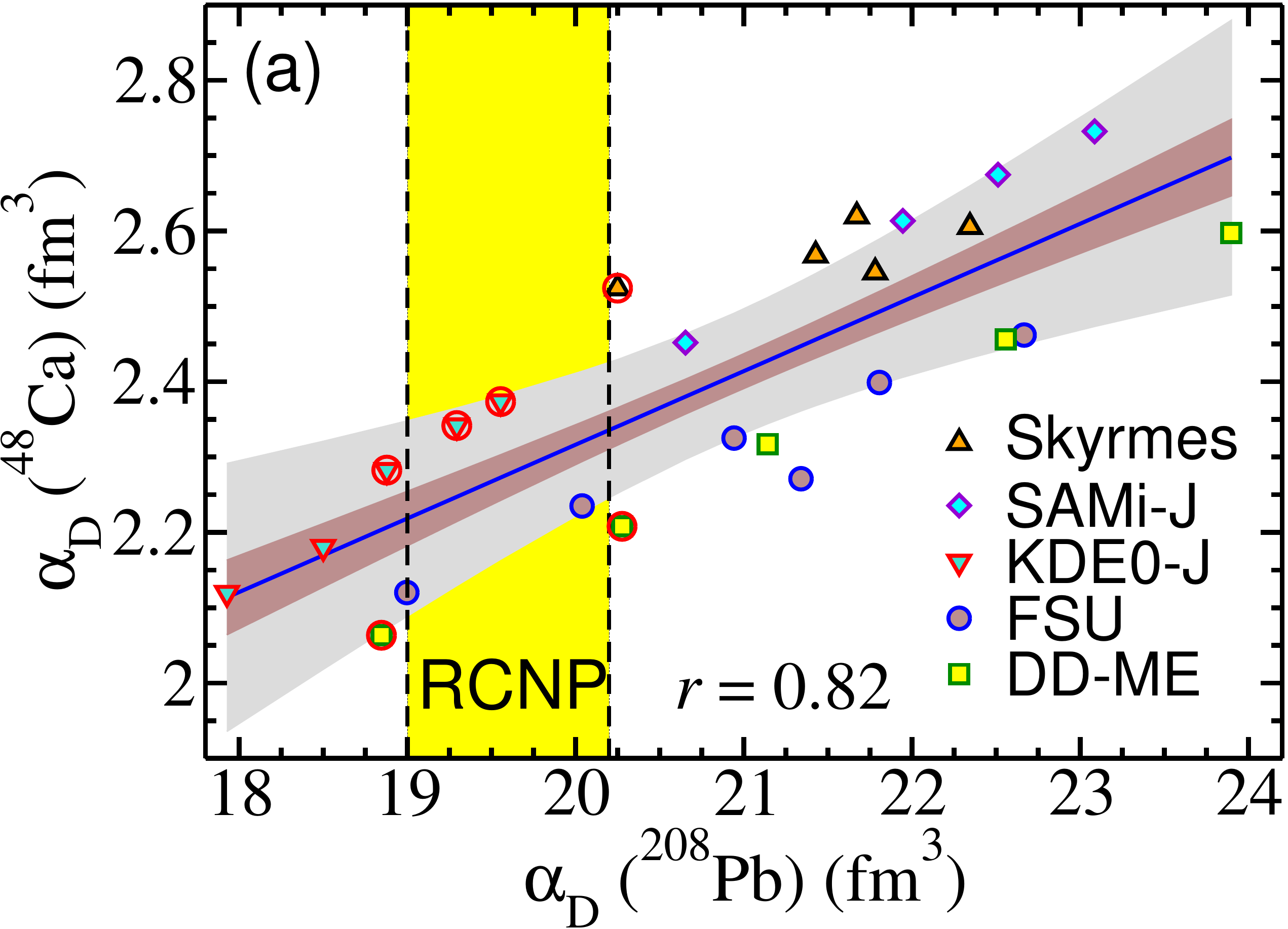}
\hspace{10pt}
\includegraphics[width=0.45\linewidth,clip=true]{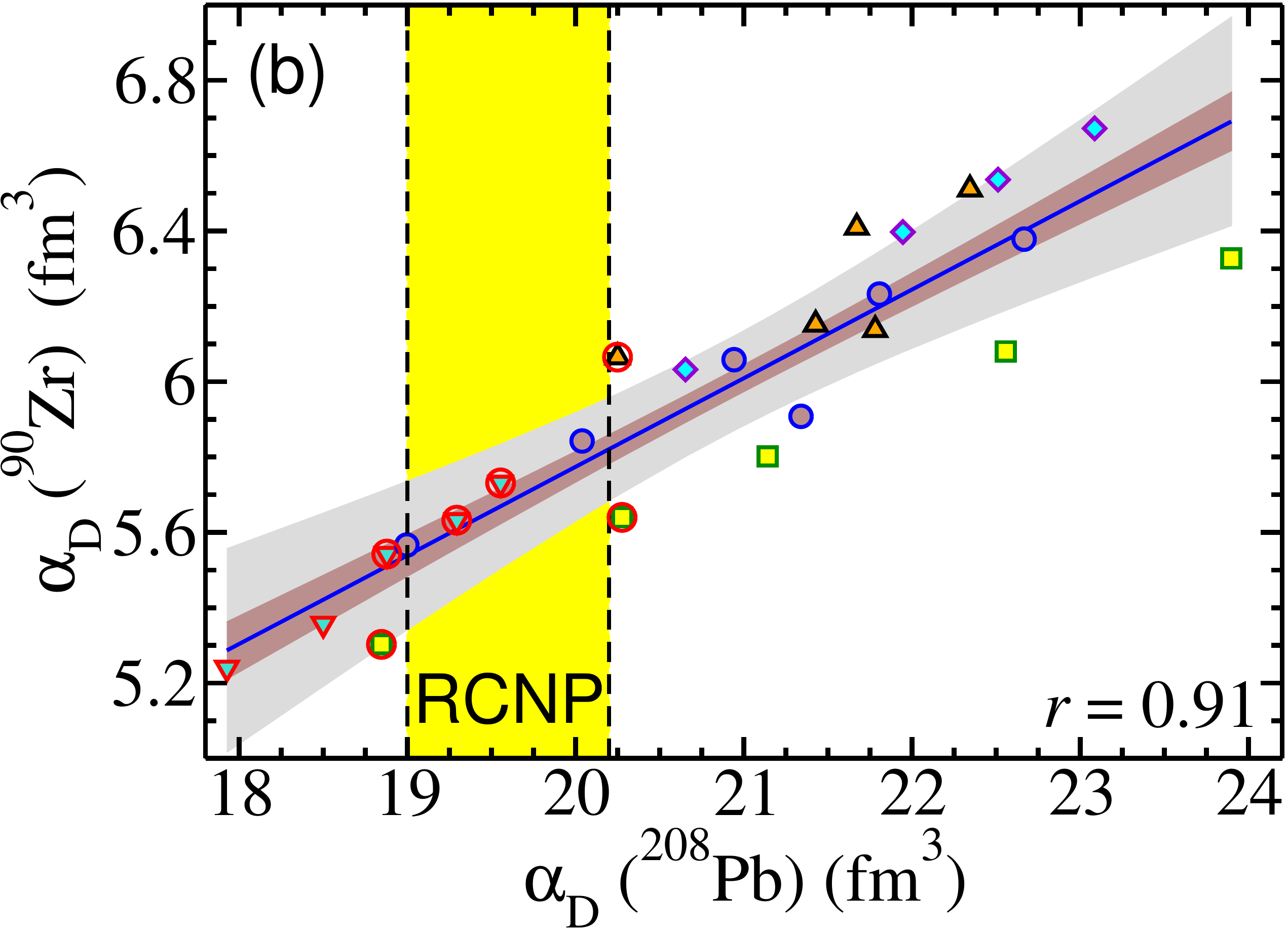}\vspace{6pt}
\includegraphics[width=0.45\linewidth,clip=true]{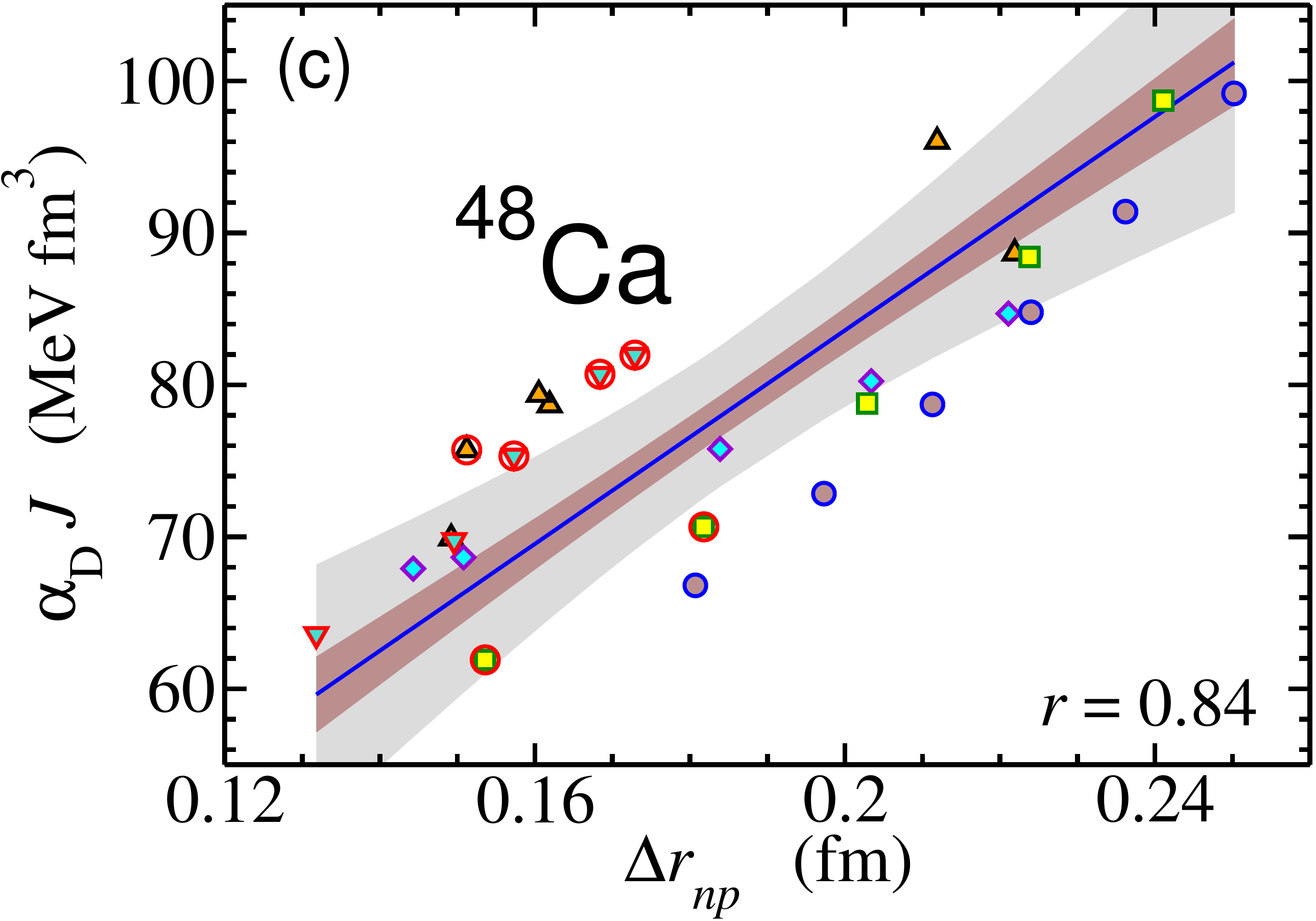}
\hspace{10pt}
\includegraphics[width=0.45\linewidth,clip=true]{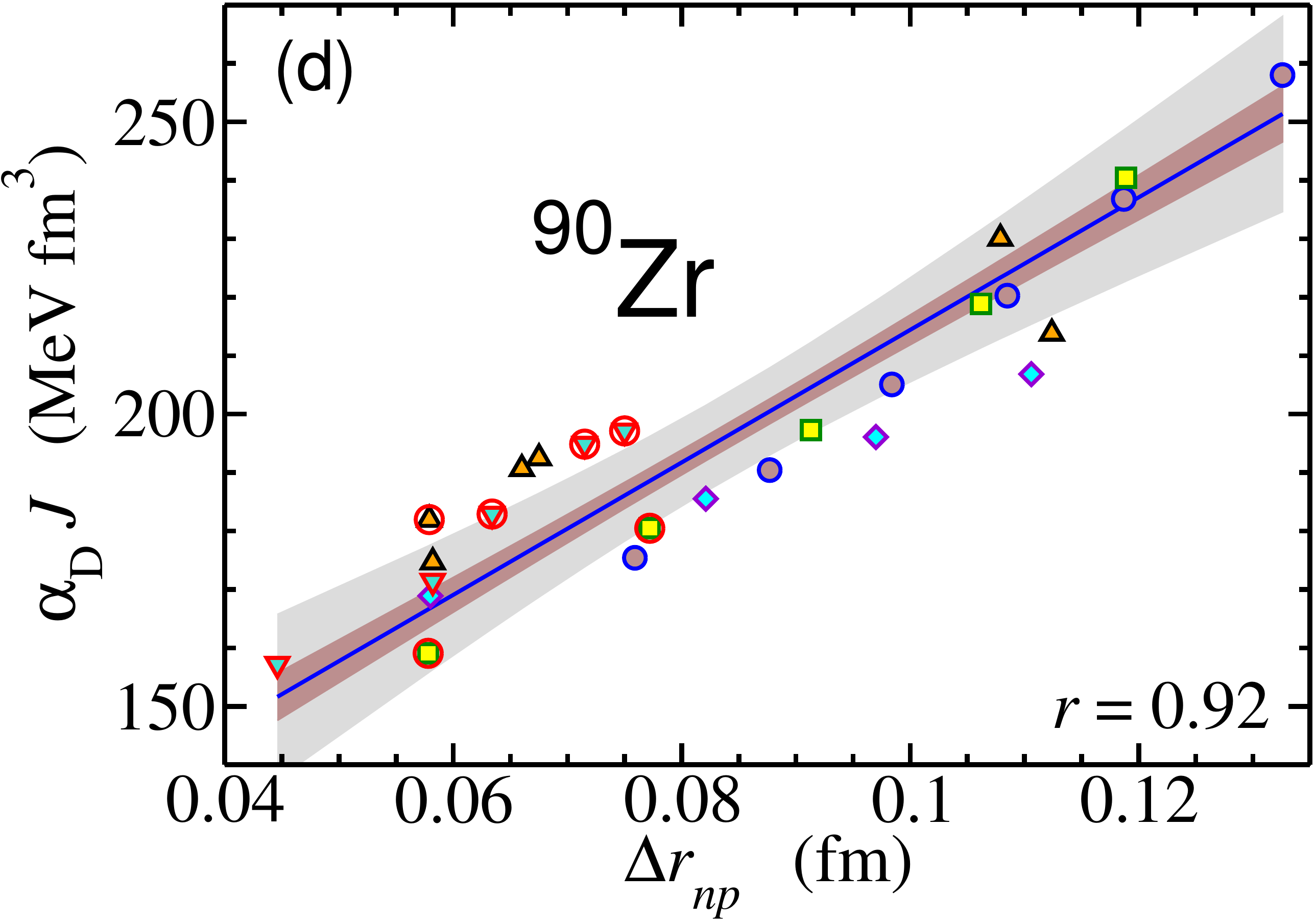}
\caption{(a) Dipole polarizability in ${}^{48}$Ca and (b) in ${}^{90}$Zr as a function of the dipole polarizability in ${}^{208}$Pb. The linear fits are (a) $\alpha_D({}^{48}$Ca)$ = (0.36\pm 0.07) + (0.10\pm 0.01)\alpha_D({}^{208}$Pb) with a correlation coefficient $r = 0.82$ and (b) $\alpha_D({}^{90}$Zr)$= (1.1\pm 0.1) + (0.24\pm 0.02)\alpha_D({}^{208}$Pb) with a correlation coefficient $r = 0.91$. (c) dipole polarizability in ${}^{48}$Ca and (d) in ${}^{90}$Zr times the symmetry energy at saturation as a function of the neutron skin thickness $\Delta r_{np}$ for the corresponding nuclei predicted by the selected EDFs. The linear fits are (c) $\alpha_D J = 12\pm 19 + (355\pm 44)\Delta r_{np}$ with a correlation coefficient $r = 0.84$, and (d) $\alpha_D J = 101\pm 26 + (1130\pm 90)\Delta r_{np}$ with a correlation coefficient $r = 0.92$. The red circles  highlight the interactions that reproduce the experimental data in ${}^{68}$Ni, ${}^{120}$Sn, and ${}^{208}$Pb. The inner (outer) colored regions depict the loci of the 95\% confidence (prediction) bands of the regression.}
\label{fig6}
\end{figure*}

\subsection{${}^{48}$Ca and ${}^{90}$Zr}
 
Experiments that measure the electric dipole polarizability of a variety of stable and unstable nuclei are carried out and being planned at RCNP and GSI. In particular, the measurement of $\alpha_D$ for both ${}^{48}$Ca and ${}^{90}$Zr is forthcoming. Hence, we now apply the technique developed in the previous section to make genuine predictions for the electric dipole polarizability, as well as the neutron skin thickness for both nuclei. Note, however, that the type of corrections discussed in Sec.\,\ref{tve} may need to be applied before comparing the measured values of the dipole polarizability to the corresponding RPA predictions.

The fact that the product of the electric dipole polarizability with the symmetry energy is better correlated to the neutron skin thickness than the polarizability alone seems to be a consistent result that has been verified in medium- and heavy-mass nuclei. However, in general, one expects that this type of correlation may weaken for light nuclei where giant resonances are usually wider and more fragmented than in heavy nuclei. This may affect moments derived from the strength distribution, such as the electric dipole polarizability.

To test this assertion we display in the upper panels of Fig.\,\ref{fig6} the correlation between $\alpha_D$ in (a) ${}^{48}$Ca and (b) ${}^{90}$Zr {\sl vs} the electric dipole polarizability in ${}^{208}$Pb for the large set of EDFs employed in this work. Similarly, the two lower panels in the figure display the $\alpha_D J$- $\Delta r_{\!np}$ correlations in (c) ${}^{48}$Ca and (d) ${}^{90}$Zr, respectively. As in the previous subsection we find that both of the upper panels display a linear correlation that may be fitted as follows: 
%%%
\begin{align}
 \!\!\alpha_D({}^{48}\textrm{Ca})\!&=\!(0.36\pm 0.07) \!+\! 
 (0.10\pm0.01)\alpha_D({}^{208}{\rm Pb}), \label{acapb} \\
 \!\!\alpha_D({}^{90}{\rm Zr})\!&=\! (1.1\phantom{0}\pm 0.1\phantom{0}) \!+\! 
(0.24\pm0.02)\alpha_D({}^{208}{\rm Pb}), \label{azrpb} 
\end{align}
%%% 
with the correlation coefficients of 0.82 for ${}^{48}$Ca and 0.91 for ${}^{90}$Zr, respectively. As in the case shown in Fig.\,\ref{fig3}, we have also calculated the ``scaled-$J$'' correlations (not plotted here) $\alpha_D({}^{48}$Ca)$J$-$\alpha_D({}^{208}$Pb)$J$ and $\alpha_D({}^{90}$Zr)$J$-$\alpha_D({}^{208}$Pb)$J$, using the same set of EDFs. We find that these correlations remain very strong even for the case of ${}^{48}$Ca, with the correlation coefficients of $r\!=\!0.94$ and $r\!=\!0.98$ for the case of Ca-Pb and Zr-Pb, respectively.

The vertical yellow band in the two upper panels of Fig.\,\ref{fig6} indicates the experimental value of the electric dipole polarizability in ${}^{208}$Pb\,\cite{tamii11}, minus the quasi-deuteron contribution. The models that lie within the interval defined by the intersection between this band (yellow) and the ``prediction band'' of the linear regression (gray area) include those models that reproduce the experimental electric dipole polarizability in ${}^{208}$Pb, and that we consider as good candidates to reproduce the corresponding quantity also in ${}^{48}$Ca and ${}^{90}$Zr. The red circles highlight the subset of models that reproduce the electric dipole polarizability in ${}^{68}$Ni, ${}^{120}$Sn, and ${}^{208}$Pb. It is remarkable that most of these models also lie within the prediction band. Further, in the two lower panels in Fig.\,\ref{fig6} that quantify the $\alpha_D J$-$\Delta r_{\!np}$ correlation, the resulting linear fits yield:
%%%
\begin{equation}
 \!\alpha_D J \!=\!
 \begin{cases}
  (\phantom{1}13\pm 19) \!+\! (\phantom{1}355\pm 44)\Delta r_{\!np}, 
  & \text{for } {}^{48}{\rm Ca}; \\
  (101\pm 26) \!+\! (1130\pm 90) \Delta r_{\!np}, 
  & \text{for } {}^{90}{\rm Zr},
 \end{cases}
 \label{fitaD2}
\end{equation}
%%%
with correlation coefficients of $0.84$ and $0.92$, respectively. Finally, using the EDFs that reproduce the experimental electric dipole polarizability in ${}^{68}$Ni, ${}^{120}$Sn and ${}^{208}$Pb, we can estimate an interval, as well as the average and standard deviation, for the polarizability and the neutron skin thickness of ${}^{48}$Ca and ${}^{90}$Zr; see Table\,\ref{table2}. 
%%%
\begin{table}[h]
\begin{center}
\begin{ruledtabular}
\begin{tabular}{cll}
Nucleus & $\hfil\Delta r_{\!np}$\,(fm) & $\hfil\alpha_{D}$\,(fm$^{3})$ \\
 \hline\rule{0pt}{2.5ex}
$^{48}$Ca  & $0.15\!-\!0.18\,(0.16\phantom{7}\pm0.01$) & 
 $2.06\!-\!2.21\,(2.3\phantom{5}\pm0.1)$ \\
$^{90}$Zr   & $0.06\!-\!0.08\,(0.067\pm0.008$) & 
 $5.30\!-\!5.64\,(5.65\pm0.23)$ 
\end{tabular} 
\caption{Estimates for the neutron skin thickness and electric dipole polarizability of $^{48}$Ca and $^{90}$Zr. Ranges as well as mean value and standard deviations are provided (see text for details). Recall that the type of corrections discussed in Sec.\,\ref{tve} may need to be applied in comparing measured values of $\alpha_{D}$ to RPA predictions.}
\label{table2} 
\end{ruledtabular}
\end{center}
\end{table}

\section{Conclusions}
\label{conclusions}

We carried out a theoretical analysis of the recently measured electric dipole polarizability in ${}^{68}$Ni, ${}^{120}$Sn, and ${}^{208}$Pb to extract information about isovector nuclear properties, such as the neutron skin thickness and the behavior of the symmetry energy around saturation density. To this end, we have computed the electric dipole polarizability of these three nuclei in a self-consistent random phase approximation using a large set of Skyrme functionals together with several families of relativistic and non-relativistic functionals. In the case of ${}^{120}$Sn, quasiparticle RPA calculations have been performed to take into account the effect of pairing correlations. Further, we have discussed in some detail how to correctly compare the measured electric dipole polarizability with theoretical results. Indeed, to directly compare the (Q)RPA results with the measured dipole polarizability it is essential to subtract the quasi-deuteron contribution\,\cite{tamii11} from the available experimental strength in both  ${}^{120}$Sn\,\cite{hashimoto15} and ${}^{208}$Pb\,\cite{tamii11}. This procedure should be systematically adopted when comparing the measured dipole polarizability with model results based on the (Q)RPA.

We have assessed, by means of (Q)RPA calculations, the validity of several correlations suggested by the Droplet-Model estimate of the electric dipole polarizability. It was found that in both ${}^{68}$Ni and ${}^{120}$Sn, the product of the electric dipole polarizability $\alpha_D$ and the symmetry energy coefficient $J$ is much better correlated with the neutron skin thickness $\Delta r_{\!np}$ than $\alpha_D$ alone. This finding is in full agreement with our previous study of the electric dipole polarizability in ${}^{208}$Pb, pointing out that this correlation is robust over the nuclear chart, with the possible exception of very light nuclei. It has also been found that while a fairly strong correlation emerges between the electric dipole polarizabilities of two neutron rich nuclei, the correlation is even stronger for the product $\alpha_DJ$. From the large set of EDFs considered in this work, we have identified a subset that simultaneously reproduces the measured electric dipole polarizability in ${}^{68}$Ni, ${}^{120}$Sn, and ${}^{208}$Pb. This subset has then been used to estimate isovector-sensitive observables, such as the neutron skin thickness and parameters of the nuclear matter symmetry energy. We estimate that the neutron skin thickness in ${}^{68}$Ni, ${}^{120}$Sn, and ${}^{208}$Pb lies in the range 0.15--0.19 fm, 0.12--0.16 fm, and 0.13--0.19 fm, respectively. The interval $30 \le\!J\!\le 35$\,MeV and $20\!\le L\!\le 66$\,MeV has been inferred for the symmetry energy $J$ and its slope at saturation density, suggesting a fairly soft symmetry energy. These estimates are consistent with other predictions of the neutron skin thickness, $J$, and $L$ extracted from various experiments that include heavy-ion collisions, giant resonances, antiprotonic atoms, hadronic probes, polarized electron scattering, as well as astrophysical observations; see, e.g., Refs.\,\cite{tsang12,lattimer2013,vinas14,Horowitz:2014bja,colo15}. Finally, the correlation between the electric dipole polarizabilities in ${}^{68}$Ni, ${}^{120}$Sn, and ${}^{208}$Pb shows that almost all the EDFs that reproduce the measured polarizability in ${}^{208}$Pb also reproduce the measured polarizabilities in ${}^{68}$Ni and ${}^{120}$Sn. This suggests the possibility of using (Q)RPA calculations to predict the presently unknown polarizability in other nuclei which could be experimentally investigated in the near future, and here is has been used to estimate the electric dipole polarizability of both ${}^{48}$Ca and ${}^{90}$Zr.

\begin{acknowledgments}
We are indebted to T. Aumann, T. Hashimoto, P. Von Neumann-Cosel, D. Rossi, and A. Tamii for useful discussions, for providing us with the full experimental set of values of the dipole polarizability in the studied nuclei and, when needed, for information about the subtraction of the so called quasi-deuteron excitations. In addition, we thank C. J. Horowitz for useful comments on the comparison between theory and experiment. M. C. and X. V. acknowledge partial support from Grant No.\  FIS2014-54672-P from Spanish MINECO and FEDER, Grant No.\ 2014SGR-401 from Generalitat de Catalunya, the Consolider-Ingenio Programme CPAN CSD2007-00042, and the project MDM-2014-0369 of ICCUB (Unidad de Excelencia Mar\'{\i}a de Maeztu) from MINECO. N.P. acknowledges partial support from FP7-PEOPLE-2011-COFUND program NEWFELPRO, the Swiss National Science Foundation, and the Croatian Science Foundation under the project Structure and Dynamics of Exotic Femtosystems (IP-2014-09-9159). J.P. acknowledges that this material is based upon work supported by the United States Department of Energy Office of Science, Office of Nuclear Physics under Award Number DE-FD05-92ER40750.
\end{acknowledgments}

%%
%=============================================================
% BIBLIOGRAPHY
%=============================================================
\bibliography{bibliography}

\begin{thebibliography}{68}%
\makeatletter
\providecommand \@ifxundefined [1]{%
 \@ifx{#1\undefined}
}%
\providecommand \@ifnum [1]{%
 \ifnum #1\expandafter \@firstoftwo
 \else \expandafter \@secondoftwo
 \fi
}%
\providecommand \@ifx [1]{%
 \ifx #1\expandafter \@firstoftwo
 \else \expandafter \@secondoftwo
 \fi
}%
\providecommand \natexlab [1]{#1}%
\providecommand \enquote  [1]{``#1''}%
\providecommand \bibnamefont  [1]{#1}%
\providecommand \bibfnamefont [1]{#1}%
\providecommand \citenamefont [1]{#1}%
\providecommand \href@noop [0]{\@secondoftwo}%
\providecommand \href [0]{\begingroup \@sanitize@url \@href}%
\providecommand \@href[1]{\@@startlink{#1}\@@href}%
\providecommand \@@href[1]{\endgroup#1\@@endlink}%
\providecommand \@sanitize@url [0]{\catcode `\\12\catcode `\$12\catcode
  `\&12\catcode `\#12\catcode `\^12\catcode `\_12\catcode `\%12\relax}%
\providecommand \@@startlink[1]{}%
\providecommand \@@endlink[0]{}%
\providecommand \url  [0]{\begingroup\@sanitize@url \@url }%
\providecommand \@url [1]{\endgroup\@href {#1}{\urlprefix }}%
\providecommand \urlprefix  [0]{URL }%
\providecommand \Eprint [0]{\href }%
\providecommand \doibase [0]{http://dx.doi.org/}%
\providecommand \selectlanguage [0]{\@gobble}%
\providecommand \bibinfo  [0]{\@secondoftwo}%
\providecommand \bibfield  [0]{\@secondoftwo}%
\providecommand \translation [1]{[#1]}%
\providecommand \BibitemOpen [0]{}%
\providecommand \bibitemStop [0]{}%
\providecommand \bibitemNoStop [0]{.\EOS\space}%
\providecommand \EOS [0]{\spacefactor3000\relax}%
\providecommand \BibitemShut  [1]{\csname bibitem#1\endcsname}%
\let\auto@bib@innerbib\@empty
%</preamble>
\bibitem [{\citenamefont {Reinhard}\ and\ \citenamefont
  {Nazarewicz}(2010)}]{reinhard10}%
  \BibitemOpen
  \bibfield  {author} {\bibinfo {author} {\bibfnamefont {P.-G.}\ \bibnamefont
  {Reinhard}}\ and\ \bibinfo {author} {\bibfnamefont {W.}~\bibnamefont
  {Nazarewicz}},\ }\href {\doibase 10.1103/PhysRevC.81.051303} {\bibfield
  {journal} {\bibinfo  {journal} {Phys. Rev. C}\ }\textbf {\bibinfo {volume}
  {81}},\ \bibinfo {pages} {051303} (\bibinfo {year} {2010})}\BibitemShut
  {NoStop}%
\bibitem [{\citenamefont {Aumann}\ and\ \citenamefont
  {Nakamura}(2013)}]{aumann13}%
  \BibitemOpen
  \bibfield  {author} {\bibinfo {author} {\bibfnamefont {T.}~\bibnamefont
  {Aumann}}\ and\ \bibinfo {author} {\bibfnamefont {T.}~\bibnamefont
  {Nakamura}},\ }\href {http://stacks.iop.org/1402-4896/2013/i=T152/a=014012}
  {\bibfield  {journal} {\bibinfo  {journal} {Physica Scripta}\ }\textbf
  {\bibinfo {volume} {2013}},\ \bibinfo {pages} {014012} (\bibinfo {year}
  {2013})}\BibitemShut {NoStop}%
\bibitem [{\citenamefont {Savran}\ \emph {et~al.}(2013)\citenamefont {Savran},
  \citenamefont {Aumann},\ and\ \citenamefont {Zilges}}]{savran2013}%
  \BibitemOpen
  \bibfield  {author} {\bibinfo {author} {\bibfnamefont {D.}~\bibnamefont
  {Savran}}, \bibinfo {author} {\bibfnamefont {T.}~\bibnamefont {Aumann}}, \
  and\ \bibinfo {author} {\bibfnamefont {A.}~\bibnamefont {Zilges}},\ }\href
  {\doibase http://dx.doi.org/10.1016/j.ppnp.2013.02.003} {\bibfield  {journal}
  {\bibinfo  {journal} {Progress in Particle and Nuclear Physics}\ }\textbf
  {\bibinfo {volume} {70}},\ \bibinfo {pages} {210 } (\bibinfo {year}
  {2013})}\BibitemShut {NoStop}%
\bibitem [{\citenamefont {Bracco}\ \emph {et~al.}(2015)\citenamefont {Bracco},
  \citenamefont {Crespi},\ and\ \citenamefont {Lanza}}]{bracco15}%
  \BibitemOpen
  \bibfield  {author} {\bibinfo {author} {\bibfnamefont {A.}~\bibnamefont
  {Bracco}}, \bibinfo {author} {\bibfnamefont {F.}~\bibnamefont {Crespi}}, \
  and\ \bibinfo {author} {\bibfnamefont {E.}~\bibnamefont {Lanza}},\ }\href
  {\doibase 10.1140/epja/i2015-15099-6} {\bibfield  {journal} {\bibinfo
  {journal} {The European Physical Journal A}\ }\textbf {\bibinfo {volume}
  {51}},\ \bibinfo {eid} {99} (\bibinfo {year} {2015}),\
  10.1140/epja/i2015-15099-6}\BibitemShut {NoStop}%
\bibitem [{\citenamefont {Paar}\ \emph {et~al.}(2007)\citenamefont {Paar},
  \citenamefont {Vretenar}, \citenamefont {Khan},\ and\ \citenamefont
  {Col\`o}}]{Paar2007}%
  \BibitemOpen
  \bibfield  {author} {\bibinfo {author} {\bibfnamefont {N.}~\bibnamefont
  {Paar}}, \bibinfo {author} {\bibfnamefont {D.}~\bibnamefont {Vretenar}},
  \bibinfo {author} {\bibfnamefont {E.}~\bibnamefont {Khan}}, \ and\ \bibinfo
  {author} {\bibfnamefont {G.}~\bibnamefont {Col\`o}},\ }\href
  {http://stacks.iop.org/0034-4885/70/i=5/a=R02} {\bibfield  {journal}
  {\bibinfo  {journal} {Reports on Progress in Physics}\ }\textbf {\bibinfo
  {volume} {70}},\ \bibinfo {pages} {691} (\bibinfo {year} {2007})}\BibitemShut
  {NoStop}%
\bibitem [{\citenamefont {Piekarewicz}(2006)}]{Piekarewicz:2006ip}%
  \BibitemOpen
  \bibfield  {author} {\bibinfo {author} {\bibfnamefont {J.}~\bibnamefont
  {Piekarewicz}},\ }\href@noop {} {\bibfield  {journal} {\bibinfo  {journal}
  {Phys. Rev.}\ }\textbf {\bibinfo {volume} {C 73}},\ \bibinfo {pages} {044325}
  (\bibinfo {year} {2006})}\BibitemShut {NoStop}%
\bibitem [{\citenamefont {Piekarewicz}(2011)}]{piekarewicz11}%
  \BibitemOpen
  \bibfield  {author} {\bibinfo {author} {\bibfnamefont {J.}~\bibnamefont
  {Piekarewicz}},\ }\href {\doibase 10.1103/PhysRevC.83.034319} {\bibfield
  {journal} {\bibinfo  {journal} {Phys. Rev. C}\ }\textbf {\bibinfo {volume}
  {83}},\ \bibinfo {pages} {034319} (\bibinfo {year} {2011})}\BibitemShut
  {NoStop}%
\bibitem [{\citenamefont {Klimkiewicz}\ \emph {et~al.}(2007)\citenamefont
  {Klimkiewicz}, \citenamefont {Paar}, \citenamefont {Adrich}, \citenamefont
  {Fallot}, \citenamefont {Boretzky}, \citenamefont {Aumann}, \citenamefont
  {Cortina-Gil}, \citenamefont {Pramanik}, \citenamefont {Elze}, \citenamefont
  {Emling}, \citenamefont {Geissel}, \citenamefont {Hellstr\"om}, \citenamefont
  {Jones}, \citenamefont {Kratz}, \citenamefont {Kulessa}, \citenamefont
  {Nociforo}, \citenamefont {Palit}, \citenamefont {Simon}, \citenamefont
  {Sur\'owka}, \citenamefont {S\"ummerer}, \citenamefont {Vretenar},\ and\
  \citenamefont {Walu\ifmmode~\acute{s}\else \'{s}\fi{}}}]{klimkiewicz07}%
  \BibitemOpen
  \bibfield  {author} {\bibinfo {author} {\bibfnamefont {A.}~\bibnamefont
  {Klimkiewicz}}, \bibinfo {author} {\bibfnamefont {N.}~\bibnamefont {Paar}},
  \bibinfo {author} {\bibfnamefont {P.}~\bibnamefont {Adrich}}, \bibinfo
  {author} {\bibfnamefont {M.}~\bibnamefont {Fallot}}, \bibinfo {author}
  {\bibfnamefont {K.}~\bibnamefont {Boretzky}}, \bibinfo {author}
  {\bibfnamefont {T.}~\bibnamefont {Aumann}}, \bibinfo {author} {\bibfnamefont
  {D.}~\bibnamefont {Cortina-Gil}}, \bibinfo {author} {\bibfnamefont {U.~D.}\
  \bibnamefont {Pramanik}}, \bibinfo {author} {\bibfnamefont {T.~W.}\
  \bibnamefont {Elze}}, \bibinfo {author} {\bibfnamefont {H.}~\bibnamefont
  {Emling}}, \bibinfo {author} {\bibfnamefont {H.}~\bibnamefont {Geissel}},
  \bibinfo {author} {\bibfnamefont {M.}~\bibnamefont {Hellstr\"om}}, \bibinfo
  {author} {\bibfnamefont {K.~L.}\ \bibnamefont {Jones}}, \bibinfo {author}
  {\bibfnamefont {J.~V.}\ \bibnamefont {Kratz}}, \bibinfo {author}
  {\bibfnamefont {R.}~\bibnamefont {Kulessa}}, \bibinfo {author} {\bibfnamefont
  {C.}~\bibnamefont {Nociforo}}, \bibinfo {author} {\bibfnamefont
  {R.}~\bibnamefont {Palit}}, \bibinfo {author} {\bibfnamefont
  {H.}~\bibnamefont {Simon}}, \bibinfo {author} {\bibfnamefont
  {G.}~\bibnamefont {Sur\'owka}}, \bibinfo {author} {\bibfnamefont
  {K.}~\bibnamefont {S\"ummerer}}, \bibinfo {author} {\bibfnamefont
  {D.}~\bibnamefont {Vretenar}}, \ and\ \bibinfo {author} {\bibfnamefont
  {W.}~\bibnamefont {Walu\ifmmode~\acute{s}\else \'{s}\fi{}}} (\bibinfo
  {collaboration} {LAND Collaboration}),\ }\href {\doibase
  10.1103/PhysRevC.76.051603} {\bibfield  {journal} {\bibinfo  {journal} {Phys.
  Rev. C}\ }\textbf {\bibinfo {volume} {76}},\ \bibinfo {pages} {051603}
  (\bibinfo {year} {2007})}\BibitemShut {NoStop}%
\bibitem [{\citenamefont {Carbone~{\it et al.}}(2010)}]{carbone10}%
  \BibitemOpen
  \bibfield  {author} {\bibinfo {author} {\bibfnamefont {A.}~\bibnamefont
  {Carbone~{\it et al.}}},\ }\href {\doibase 10.1103/PhysRevC.81.041301}
  {\bibfield  {journal} {\bibinfo  {journal} {Phys. Rev. C}\ }\textbf {\bibinfo
  {volume} {81}},\ \bibinfo {pages} {041301} (\bibinfo {year}
  {2010})}\BibitemShut {NoStop}%
\bibitem [{\citenamefont {Roca-Maza}\ \emph
  {et~al.}(2012{\natexlab{a}})\citenamefont {Roca-Maza}, \citenamefont {Pozzi},
  \citenamefont {Brenna}, \citenamefont {Mizuyama},\ and\ \citenamefont
  {Col\`o}}]{roca-maza12a}%
  \BibitemOpen
  \bibfield  {author} {\bibinfo {author} {\bibfnamefont {X.}~\bibnamefont
  {Roca-Maza}}, \bibinfo {author} {\bibfnamefont {G.}~\bibnamefont {Pozzi}},
  \bibinfo {author} {\bibfnamefont {M.}~\bibnamefont {Brenna}}, \bibinfo
  {author} {\bibfnamefont {K.}~\bibnamefont {Mizuyama}}, \ and\ \bibinfo
  {author} {\bibfnamefont {G.}~\bibnamefont {Col\`o}},\ }\href {\doibase
  10.1103/PhysRevC.85.024601} {\bibfield  {journal} {\bibinfo  {journal} {Phys.
  Rev. C}\ }\textbf {\bibinfo {volume} {85}},\ \bibinfo {pages} {024601}
  (\bibinfo {year} {2012}{\natexlab{a}})}\BibitemShut {NoStop}%
\bibitem [{\citenamefont {Baran}\ \emph {et~al.}(2015)\citenamefont {Baran},
  \citenamefont {Colonna}, \citenamefont {Di~Toro}, \citenamefont {Croitoru},\
  and\ \citenamefont {Nicolin}}]{baran15}%
  \BibitemOpen
  \bibfield  {author} {\bibinfo {author} {\bibfnamefont {V.}~\bibnamefont
  {Baran}}, \bibinfo {author} {\bibfnamefont {M.}~\bibnamefont {Colonna}},
  \bibinfo {author} {\bibfnamefont {M.}~\bibnamefont {Di~Toro}}, \bibinfo
  {author} {\bibfnamefont {A.}~\bibnamefont {Croitoru}}, \ and\ \bibinfo
  {author} {\bibfnamefont {A.~I.}\ \bibnamefont {Nicolin}},\ }\href
  {http://scitation.aip.org/content/aip/proceeding/aipcp/10.1063/1.4909584",
  doi = "http://dx.doi.org/10.1063/1.4909584} {\bibfield  {journal} {\bibinfo
  {journal} {AIP Conference Proceedings}\ }\textbf {\bibinfo {volume} {1645}},\
  \bibinfo {pages} {267} (\bibinfo {year} {2015})}\BibitemShut {NoStop}%
\bibitem [{\citenamefont {{Papakonstantinou}}\ \emph
  {et~al.}(2015)\citenamefont {{Papakonstantinou}}, \citenamefont {{Hergert}},\
  and\ \citenamefont {{Roth}}}]{papakonstantinou15}%
  \BibitemOpen
  \bibfield  {author} {\bibinfo {author} {\bibfnamefont {P.}~\bibnamefont
  {{Papakonstantinou}}}, \bibinfo {author} {\bibfnamefont {H.}~\bibnamefont
  {{Hergert}}}, \ and\ \bibinfo {author} {\bibfnamefont {R.}~\bibnamefont
  {{Roth}}},\ }\href@noop {} {\bibfield  {journal} {\bibinfo  {journal} {ArXiv
  e-prints}\ } (\bibinfo {year} {2015})},\ \Eprint
  {http://arxiv.org/abs/1506.04841} {arXiv:1506.04841 [nucl-th]} \BibitemShut
  {NoStop}%
\bibitem [{\citenamefont {Tamii~{\it et al.}}(2011)}]{tamii11}%
  \BibitemOpen
  \bibfield  {author} {\bibinfo {author} {\bibfnamefont {A.}~\bibnamefont
  {Tamii~{\it et al.}}},\ }\href {\doibase 10.1103/PhysRevLett.107.062502}
  {\bibfield  {journal} {\bibinfo  {journal} {Phys. Rev. Lett.}\ }\textbf
  {\bibinfo {volume} {107}},\ \bibinfo {pages} {062502} (\bibinfo {year}
  {2011})}\BibitemShut {NoStop}%
\bibitem [{\citenamefont {Schelhaas}\ \emph {et~al.}(1988)\citenamefont
  {Schelhaas}, \citenamefont {Henneberg}, \citenamefont {Sanzone-Arenhövel},
  \citenamefont {Wieloch-Laufenberg}, \citenamefont {Zurmühl}, \citenamefont
  {Ziegler}, \citenamefont {Schumacher},\ and\ \citenamefont
  {Wolf}}]{schelhaas1988}%
  \BibitemOpen
  \bibfield  {author} {\bibinfo {author} {\bibfnamefont {K.}~\bibnamefont
  {Schelhaas}}, \bibinfo {author} {\bibfnamefont {J.}~\bibnamefont
  {Henneberg}}, \bibinfo {author} {\bibfnamefont {M.}~\bibnamefont
  {Sanzone-Arenhövel}}, \bibinfo {author} {\bibfnamefont {N.}~\bibnamefont
  {Wieloch-Laufenberg}}, \bibinfo {author} {\bibfnamefont {U.}~\bibnamefont
  {Zurmühl}}, \bibinfo {author} {\bibfnamefont {B.}~\bibnamefont {Ziegler}},
  \bibinfo {author} {\bibfnamefont {M.}~\bibnamefont {Schumacher}}, \ and\
  \bibinfo {author} {\bibfnamefont {F.}~\bibnamefont {Wolf}},\ }\href {\doibase
  http://dx.doi.org/10.1016/0375-9474(88)90149-2} {\bibfield  {journal}
  {\bibinfo  {journal} {Nuclear Physics A}\ }\textbf {\bibinfo {volume}
  {489}},\ \bibinfo {pages} {189 } (\bibinfo {year} {1988})}\BibitemShut
  {NoStop}%
\bibitem [{\citenamefont {Veyssiere}\ \emph {et~al.}(1970)\citenamefont
  {Veyssiere}, \citenamefont {Beil}, \citenamefont {Bergere}, \citenamefont
  {Carlos},\ and\ \citenamefont {Lepretre}}]{veyssiere1970}%
  \BibitemOpen
  \bibfield  {author} {\bibinfo {author} {\bibfnamefont {A.}~\bibnamefont
  {Veyssiere}}, \bibinfo {author} {\bibfnamefont {H.}~\bibnamefont {Beil}},
  \bibinfo {author} {\bibfnamefont {R.}~\bibnamefont {Bergere}}, \bibinfo
  {author} {\bibfnamefont {P.}~\bibnamefont {Carlos}}, \ and\ \bibinfo {author}
  {\bibfnamefont {A.}~\bibnamefont {Lepretre}},\ }\href {\doibase
  http://dx.doi.org/10.1016/0375-9474(70)90727-X} {\bibfield  {journal}
  {\bibinfo  {journal} {Nuclear Physics A}\ }\textbf {\bibinfo {volume}
  {159}},\ \bibinfo {pages} {561 } (\bibinfo {year} {1970})}\BibitemShut
  {NoStop}%
\bibitem [{\citenamefont {Hashimoto}\ \emph {et~al.}(2015)\citenamefont
  {Hashimoto}, \citenamefont {Krumbholz}, \citenamefont {Reinhard},
  \citenamefont {Tamii}, \citenamefont {von Neumann-Cosel}, \citenamefont
  {Adachi}, \citenamefont {Aoi}, \citenamefont {Bertulani}, \citenamefont
  {Fujita}, \citenamefont {Fujita}, \citenamefont {Ganio\ifmmode~\check{g}\else
  \v{g}\fi{}lu}, \citenamefont {Hatanaka}, \citenamefont {Ideguchi},
  \citenamefont {Iwamoto}, \citenamefont {Kawabata}, \citenamefont {Khai},
  \citenamefont {Krugmann}, \citenamefont {Martin}, \citenamefont {Matsubara},
  \citenamefont {Miki}, \citenamefont {Neveling}, \citenamefont {Okamura},
  \citenamefont {Ong}, \citenamefont {Poltoratska}, \citenamefont {Ponomarev},
  \citenamefont {Richter}, \citenamefont {Sakaguchi}, \citenamefont {Shimbara},
  \citenamefont {Shimizu}, \citenamefont {Simonis}, \citenamefont {Smit},
  \citenamefont {S\"usoy}, \citenamefont {Suzuki}, \citenamefont {Thies},
  \citenamefont {Yosoi},\ and\ \citenamefont {Zenihiro}}]{hashimoto15}%
  \BibitemOpen
  \bibfield  {author} {\bibinfo {author} {\bibfnamefont {T.}~\bibnamefont
  {Hashimoto}}, \bibinfo {author} {\bibfnamefont {A.~M.}\ \bibnamefont
  {Krumbholz}}, \bibinfo {author} {\bibfnamefont {P.-G.}\ \bibnamefont
  {Reinhard}}, \bibinfo {author} {\bibfnamefont {A.}~\bibnamefont {Tamii}},
  \bibinfo {author} {\bibfnamefont {P.}~\bibnamefont {von Neumann-Cosel}},
  \bibinfo {author} {\bibfnamefont {T.}~\bibnamefont {Adachi}}, \bibinfo
  {author} {\bibfnamefont {N.}~\bibnamefont {Aoi}}, \bibinfo {author}
  {\bibfnamefont {C.~A.}\ \bibnamefont {Bertulani}}, \bibinfo {author}
  {\bibfnamefont {H.}~\bibnamefont {Fujita}}, \bibinfo {author} {\bibfnamefont
  {Y.}~\bibnamefont {Fujita}}, \bibinfo {author} {\bibfnamefont
  {E.}~\bibnamefont {Ganio\ifmmode~\check{g}\else \v{g}\fi{}lu}}, \bibinfo
  {author} {\bibfnamefont {K.}~\bibnamefont {Hatanaka}}, \bibinfo {author}
  {\bibfnamefont {E.}~\bibnamefont {Ideguchi}}, \bibinfo {author}
  {\bibfnamefont {C.}~\bibnamefont {Iwamoto}}, \bibinfo {author} {\bibfnamefont
  {T.}~\bibnamefont {Kawabata}}, \bibinfo {author} {\bibfnamefont {N.~T.}\
  \bibnamefont {Khai}}, \bibinfo {author} {\bibfnamefont {A.}~\bibnamefont
  {Krugmann}}, \bibinfo {author} {\bibfnamefont {D.}~\bibnamefont {Martin}},
  \bibinfo {author} {\bibfnamefont {H.}~\bibnamefont {Matsubara}}, \bibinfo
  {author} {\bibfnamefont {K.}~\bibnamefont {Miki}}, \bibinfo {author}
  {\bibfnamefont {R.}~\bibnamefont {Neveling}}, \bibinfo {author}
  {\bibfnamefont {H.}~\bibnamefont {Okamura}}, \bibinfo {author} {\bibfnamefont
  {H.~J.}\ \bibnamefont {Ong}}, \bibinfo {author} {\bibfnamefont
  {I.}~\bibnamefont {Poltoratska}}, \bibinfo {author} {\bibfnamefont {V.~Y.}\
  \bibnamefont {Ponomarev}}, \bibinfo {author} {\bibfnamefont {A.}~\bibnamefont
  {Richter}}, \bibinfo {author} {\bibfnamefont {H.}~\bibnamefont {Sakaguchi}},
  \bibinfo {author} {\bibfnamefont {Y.}~\bibnamefont {Shimbara}}, \bibinfo
  {author} {\bibfnamefont {Y.}~\bibnamefont {Shimizu}}, \bibinfo {author}
  {\bibfnamefont {J.}~\bibnamefont {Simonis}}, \bibinfo {author} {\bibfnamefont
  {F.~D.}\ \bibnamefont {Smit}}, \bibinfo {author} {\bibfnamefont
  {G.}~\bibnamefont {S\"usoy}}, \bibinfo {author} {\bibfnamefont
  {T.}~\bibnamefont {Suzuki}}, \bibinfo {author} {\bibfnamefont {J.~H.}\
  \bibnamefont {Thies}}, \bibinfo {author} {\bibfnamefont {M.}~\bibnamefont
  {Yosoi}}, \ and\ \bibinfo {author} {\bibfnamefont {J.}~\bibnamefont
  {Zenihiro}},\ }\href {\doibase 10.1103/PhysRevC.92.031305} {\bibfield
  {journal} {\bibinfo  {journal} {Phys. Rev. C}\ }\textbf {\bibinfo {volume}
  {92}},\ \bibinfo {pages} {031305} (\bibinfo {year} {2015})}\BibitemShut
  {NoStop}%
\bibitem [{\citenamefont {Lepretre}\ \emph {et~al.}(1981)\citenamefont
  {Lepretre}, \citenamefont {Beil}, \citenamefont {Bergère}, \citenamefont
  {Carlos}, \citenamefont {Fagot}, \citenamefont {Miniac},\ and\ \citenamefont
  {Veyssière}}]{lepretre1981}%
  \BibitemOpen
  \bibfield  {author} {\bibinfo {author} {\bibfnamefont {A.}~\bibnamefont
  {Lepretre}}, \bibinfo {author} {\bibfnamefont {H.}~\bibnamefont {Beil}},
  \bibinfo {author} {\bibfnamefont {R.}~\bibnamefont {Bergère}}, \bibinfo
  {author} {\bibfnamefont {P.}~\bibnamefont {Carlos}}, \bibinfo {author}
  {\bibfnamefont {J.}~\bibnamefont {Fagot}}, \bibinfo {author} {\bibfnamefont
  {A.~D.}\ \bibnamefont {Miniac}}, \ and\ \bibinfo {author} {\bibfnamefont
  {A.}~\bibnamefont {Veyssière}},\ }\href {\doibase
  http://dx.doi.org/10.1016/0375-9474(81)90516-9} {\bibfield  {journal}
  {\bibinfo  {journal} {Nuclear Physics A}\ }\textbf {\bibinfo {volume}
  {367}},\ \bibinfo {pages} {237 } (\bibinfo {year} {1981})}\BibitemShut
  {NoStop}%
\bibitem [{\citenamefont {Wieland}\ \emph {et~al.}(2009)\citenamefont {Wieland}
  \emph {et~al.}}]{Wieland:2009}%
  \BibitemOpen
  \bibfield  {author} {\bibinfo {author} {\bibfnamefont {O.}~\bibnamefont
  {Wieland}} \emph {et~al.},\ }\href {\doibase 10.1103/PhysRevLett.102.092502}
  {\bibfield  {journal} {\bibinfo  {journal} {Phys. Rev. Lett.}\ }\textbf
  {\bibinfo {volume} {102}},\ \bibinfo {pages} {092502} (\bibinfo {year}
  {2009})}\BibitemShut {NoStop}%
\bibitem [{\citenamefont {Rossi}\ \emph {et~al.}(2013)\citenamefont {Rossi},
  \citenamefont {Adrich}, \citenamefont {Aksouh}, \citenamefont {Alvarez-Pol},
  \citenamefont {Aumann}, \citenamefont {Benlliure}, \citenamefont {B\"ohmer},
  \citenamefont {Boretzky}, \citenamefont {Casarejos}, \citenamefont
  {Chartier}, \citenamefont {Chatillon}, \citenamefont {Cortina-Gil},
  \citenamefont {Datta~Pramanik}, \citenamefont {Emling}, \citenamefont
  {Ershova}, \citenamefont {Fernandez-Dominguez}, \citenamefont {Geissel},
  \citenamefont {Gorska}, \citenamefont {Heil}, \citenamefont {Johansson},
  \citenamefont {Junghans}, \citenamefont {Kelic-Heil}, \citenamefont
  {Kiselev}, \citenamefont {Klimkiewicz}, \citenamefont {Kratz}, \citenamefont
  {Kr\"ucken}, \citenamefont {Kurz}, \citenamefont {Labiche}, \citenamefont
  {Le~Bleis}, \citenamefont {Lemmon}, \citenamefont {Litvinov}, \citenamefont
  {Mahata}, \citenamefont {Maierbeck}, \citenamefont {Movsesyan}, \citenamefont
  {Nilsson}, \citenamefont {Nociforo}, \citenamefont {Palit}, \citenamefont
  {Paschalis}, \citenamefont {Plag}, \citenamefont {Reifarth}, \citenamefont
  {Savran}, \citenamefont {Scheit}, \citenamefont {Simon}, \citenamefont
  {S\"ummerer}, \citenamefont {Wagner}, \citenamefont
  {Walu\ifmmode~\acute{s}\else \'{s}\fi{}}, \citenamefont {Weick},\ and\
  \citenamefont {Winkler}}]{rossi13}%
  \BibitemOpen
  \bibfield  {author} {\bibinfo {author} {\bibfnamefont {D.~M.}\ \bibnamefont
  {Rossi}}, \bibinfo {author} {\bibfnamefont {P.}~\bibnamefont {Adrich}},
  \bibinfo {author} {\bibfnamefont {F.}~\bibnamefont {Aksouh}}, \bibinfo
  {author} {\bibfnamefont {H.}~\bibnamefont {Alvarez-Pol}}, \bibinfo {author}
  {\bibfnamefont {T.}~\bibnamefont {Aumann}}, \bibinfo {author} {\bibfnamefont
  {J.}~\bibnamefont {Benlliure}}, \bibinfo {author} {\bibfnamefont
  {M.}~\bibnamefont {B\"ohmer}}, \bibinfo {author} {\bibfnamefont
  {K.}~\bibnamefont {Boretzky}}, \bibinfo {author} {\bibfnamefont
  {E.}~\bibnamefont {Casarejos}}, \bibinfo {author} {\bibfnamefont
  {M.}~\bibnamefont {Chartier}}, \bibinfo {author} {\bibfnamefont
  {A.}~\bibnamefont {Chatillon}}, \bibinfo {author} {\bibfnamefont
  {D.}~\bibnamefont {Cortina-Gil}}, \bibinfo {author} {\bibfnamefont
  {U.}~\bibnamefont {Datta~Pramanik}}, \bibinfo {author} {\bibfnamefont
  {H.}~\bibnamefont {Emling}}, \bibinfo {author} {\bibfnamefont
  {O.}~\bibnamefont {Ershova}}, \bibinfo {author} {\bibfnamefont
  {B.}~\bibnamefont {Fernandez-Dominguez}}, \bibinfo {author} {\bibfnamefont
  {H.}~\bibnamefont {Geissel}}, \bibinfo {author} {\bibfnamefont
  {M.}~\bibnamefont {Gorska}}, \bibinfo {author} {\bibfnamefont
  {M.}~\bibnamefont {Heil}}, \bibinfo {author} {\bibfnamefont {H.~T.}\
  \bibnamefont {Johansson}}, \bibinfo {author} {\bibfnamefont {A.}~\bibnamefont
  {Junghans}}, \bibinfo {author} {\bibfnamefont {A.}~\bibnamefont
  {Kelic-Heil}}, \bibinfo {author} {\bibfnamefont {O.}~\bibnamefont {Kiselev}},
  \bibinfo {author} {\bibfnamefont {A.}~\bibnamefont {Klimkiewicz}}, \bibinfo
  {author} {\bibfnamefont {J.~V.}\ \bibnamefont {Kratz}}, \bibinfo {author}
  {\bibfnamefont {R.}~\bibnamefont {Kr\"ucken}}, \bibinfo {author}
  {\bibfnamefont {N.}~\bibnamefont {Kurz}}, \bibinfo {author} {\bibfnamefont
  {M.}~\bibnamefont {Labiche}}, \bibinfo {author} {\bibfnamefont
  {T.}~\bibnamefont {Le~Bleis}}, \bibinfo {author} {\bibfnamefont
  {R.}~\bibnamefont {Lemmon}}, \bibinfo {author} {\bibfnamefont {Y.~A.}\
  \bibnamefont {Litvinov}}, \bibinfo {author} {\bibfnamefont {K.}~\bibnamefont
  {Mahata}}, \bibinfo {author} {\bibfnamefont {P.}~\bibnamefont {Maierbeck}},
  \bibinfo {author} {\bibfnamefont {A.}~\bibnamefont {Movsesyan}}, \bibinfo
  {author} {\bibfnamefont {T.}~\bibnamefont {Nilsson}}, \bibinfo {author}
  {\bibfnamefont {C.}~\bibnamefont {Nociforo}}, \bibinfo {author}
  {\bibfnamefont {R.}~\bibnamefont {Palit}}, \bibinfo {author} {\bibfnamefont
  {S.}~\bibnamefont {Paschalis}}, \bibinfo {author} {\bibfnamefont
  {R.}~\bibnamefont {Plag}}, \bibinfo {author} {\bibfnamefont {R.}~\bibnamefont
  {Reifarth}}, \bibinfo {author} {\bibfnamefont {D.}~\bibnamefont {Savran}},
  \bibinfo {author} {\bibfnamefont {H.}~\bibnamefont {Scheit}}, \bibinfo
  {author} {\bibfnamefont {H.}~\bibnamefont {Simon}}, \bibinfo {author}
  {\bibfnamefont {K.}~\bibnamefont {S\"ummerer}}, \bibinfo {author}
  {\bibfnamefont {A.}~\bibnamefont {Wagner}}, \bibinfo {author} {\bibfnamefont
  {W.}~\bibnamefont {Walu\ifmmode~\acute{s}\else \'{s}\fi{}}}, \bibinfo
  {author} {\bibfnamefont {H.}~\bibnamefont {Weick}}, \ and\ \bibinfo {author}
  {\bibfnamefont {M.}~\bibnamefont {Winkler}},\ }\href {\doibase
  10.1103/PhysRevLett.111.242503} {\bibfield  {journal} {\bibinfo  {journal}
  {Phys. Rev. Lett.}\ }\textbf {\bibinfo {volume} {111}},\ \bibinfo {pages}
  {242503} (\bibinfo {year} {2013})}\BibitemShut {NoStop}%
\bibitem [{\citenamefont {Piekarewicz~{\it et al.}}(2012)}]{piekarewicz12}%
  \BibitemOpen
  \bibfield  {author} {\bibinfo {author} {\bibfnamefont {J.}~\bibnamefont
  {Piekarewicz~{\it et al.}}},\ }\href {\doibase 10.1103/PhysRevC.85.041302}
  {\bibfield  {journal} {\bibinfo  {journal} {Phys. Rev. C}\ }\textbf {\bibinfo
  {volume} {85}},\ \bibinfo {pages} {041302} (\bibinfo {year}
  {2012})}\BibitemShut {NoStop}%
\bibitem [{\citenamefont {Roca-Maza}\ \emph {et~al.}(2013)\citenamefont
  {Roca-Maza}, \citenamefont {Brenna}, \citenamefont {Col\`o}, \citenamefont
  {Centelles}, \citenamefont {Vi\~nas}, \citenamefont {Agrawal}, \citenamefont
  {Paar}, \citenamefont {Vretenar},\ and\ \citenamefont
  {Piekarewicz}}]{roca-maza13a}%
  \BibitemOpen
  \bibfield  {author} {\bibinfo {author} {\bibfnamefont {X.}~\bibnamefont
  {Roca-Maza}}, \bibinfo {author} {\bibfnamefont {M.}~\bibnamefont {Brenna}},
  \bibinfo {author} {\bibfnamefont {G.}~\bibnamefont {Col\`o}}, \bibinfo
  {author} {\bibfnamefont {M.}~\bibnamefont {Centelles}}, \bibinfo {author}
  {\bibfnamefont {X.}~\bibnamefont {Vi\~nas}}, \bibinfo {author} {\bibfnamefont
  {B.~K.}\ \bibnamefont {Agrawal}}, \bibinfo {author} {\bibfnamefont
  {N.}~\bibnamefont {Paar}}, \bibinfo {author} {\bibfnamefont {D.}~\bibnamefont
  {Vretenar}}, \ and\ \bibinfo {author} {\bibfnamefont {J.}~\bibnamefont
  {Piekarewicz}},\ }\href {\doibase 10.1103/PhysRevC.88.024316} {\bibfield
  {journal} {\bibinfo  {journal} {Phys. Rev. C}\ }\textbf {\bibinfo {volume}
  {88}},\ \bibinfo {pages} {024316} (\bibinfo {year} {2013})}\BibitemShut
  {NoStop}%
\bibitem [{\citenamefont {Ring}\ and\ \citenamefont
  {Schuck}(2004)}]{ringschuck}%
  \BibitemOpen
  \bibfield  {author} {\bibinfo {author} {\bibfnamefont {P.}~\bibnamefont
  {Ring}}\ and\ \bibinfo {author} {\bibfnamefont {P.}~\bibnamefont {Schuck}},\
  }\enquote {\bibinfo {title} {The nuclear many-body problem},}\ \ (\bibinfo
  {publisher} {Springer},\ \bibinfo {year} {2004})\BibitemShut {NoStop}%
\bibitem [{ats()}]{atsushi}%
  \BibitemOpen
  \href@noop {} {}\bibinfo {note} {Private communication with A.
  Tamii.}\BibitemShut {Stop}%
\bibitem [{\citenamefont {Fracasso}\ and\ \citenamefont
  {Col\`o}(2005)}]{fracasso2005}%
  \BibitemOpen
  \bibfield  {author} {\bibinfo {author} {\bibfnamefont {S.}~\bibnamefont
  {Fracasso}}\ and\ \bibinfo {author} {\bibfnamefont {G.}~\bibnamefont
  {Col\`o}},\ }\href {\doibase 10.1103/PhysRevC.72.064310} {\bibfield
  {journal} {\bibinfo  {journal} {Phys. Rev. C}\ }\textbf {\bibinfo {volume}
  {72}},\ \bibinfo {pages} {064310} (\bibinfo {year} {2005})}\BibitemShut
  {NoStop}%
\bibitem [{\citenamefont {Berger}\ \emph {et~al.}(1991)\citenamefont {Berger},
  \citenamefont {Girod},\ and\ \citenamefont {Gogny}}]{berger91}%
  \BibitemOpen
  \bibfield  {author} {\bibinfo {author} {\bibfnamefont {J.}~\bibnamefont
  {Berger}}, \bibinfo {author} {\bibfnamefont {M.}~\bibnamefont {Girod}}, \
  and\ \bibinfo {author} {\bibfnamefont {D.}~\bibnamefont {Gogny}},\ }\href
  {\doibase http://dx.doi.org/10.1016/0010-4655(91)90263-K} {\bibfield
  {journal} {\bibinfo  {journal} {Computer Physics Communications}\ }\textbf
  {\bibinfo {volume} {63}},\ \bibinfo {pages} {365 } (\bibinfo {year}
  {1991})}\BibitemShut {NoStop}%
\bibitem [{\citenamefont {Col\`o}\ \emph {et~al.}(2013)\citenamefont {Col\`o},
  \citenamefont {Cao}, \citenamefont {Giai},\ and\ \citenamefont
  {Capelli}}]{colo13}%
  \BibitemOpen
  \bibfield  {author} {\bibinfo {author} {\bibfnamefont {G.}~\bibnamefont
  {Col\`o}}, \bibinfo {author} {\bibfnamefont {L.}~\bibnamefont {Cao}},
  \bibinfo {author} {\bibfnamefont {N.~V.}\ \bibnamefont {Giai}}, \ and\
  \bibinfo {author} {\bibfnamefont {L.}~\bibnamefont {Capelli}},\ }\href
  {\doibase 10.1016/j.cpc.2012.07.016} {\bibfield  {journal} {\bibinfo
  {journal} {Computer Physics Communications}\ }\textbf {\bibinfo {volume}
  {184}},\ \bibinfo {pages} {142 } (\bibinfo {year} {2013})}\BibitemShut
  {NoStop}%
\bibitem [{\citenamefont {Lalazissis}\ \emph {et~al.}(2005)\citenamefont
  {Lalazissis}, \citenamefont {Nik\ifmmode \check{s}\else
  \v{s}\fi{}i\ifmmode~\acute{c}\else \'{c}\fi{}}, \citenamefont {Vretenar},\
  and\ \citenamefont {Ring}}]{ddme2}%
  \BibitemOpen
  \bibfield  {author} {\bibinfo {author} {\bibfnamefont {G.~A.}\ \bibnamefont
  {Lalazissis}}, \bibinfo {author} {\bibfnamefont {T.}~\bibnamefont
  {Nik\ifmmode \check{s}\else \v{s}\fi{}i\ifmmode~\acute{c}\else \'{c}\fi{}}},
  \bibinfo {author} {\bibfnamefont {D.}~\bibnamefont {Vretenar}}, \ and\
  \bibinfo {author} {\bibfnamefont {P.}~\bibnamefont {Ring}},\ }\href {\doibase
  10.1103/PhysRevC.71.024312} {\bibfield  {journal} {\bibinfo  {journal} {Phys.
  Rev. C}\ }\textbf {\bibinfo {volume} {71}},\ \bibinfo {pages} {024312}
  (\bibinfo {year} {2005})}\BibitemShut {NoStop}%
\bibitem [{\citenamefont {Vretenar}\ \emph {et~al.}(2003)\citenamefont
  {Vretenar}, \citenamefont {Nik\ifmmode \check{s}\else
  \v{s}\fi{}i\ifmmode~\acute{c}\else \'{c}\fi{}},\ and\ \citenamefont
  {Ring}}]{ddme}%
  \BibitemOpen
  \bibfield  {author} {\bibinfo {author} {\bibfnamefont {D.}~\bibnamefont
  {Vretenar}}, \bibinfo {author} {\bibfnamefont {T.}~\bibnamefont {Nik\ifmmode
  \check{s}\else \v{s}\fi{}i\ifmmode~\acute{c}\else \'{c}\fi{}}}, \ and\
  \bibinfo {author} {\bibfnamefont {P.}~\bibnamefont {Ring}},\ }\href {\doibase
  10.1103/PhysRevC.68.024310} {\bibfield  {journal} {\bibinfo  {journal} {Phys.
  Rev. C}\ }\textbf {\bibinfo {volume} {68}},\ \bibinfo {pages} {024310}
  (\bibinfo {year} {2003})}\BibitemShut {NoStop}%
\bibitem [{\citenamefont {Bohigas}\ \emph {et~al.}(1979)\citenamefont
  {Bohigas}, \citenamefont {Lane},\ and\ \citenamefont
  {Martorell}}]{bohigas1979}%
  \BibitemOpen
  \bibfield  {author} {\bibinfo {author} {\bibfnamefont {O.}~\bibnamefont
  {Bohigas}}, \bibinfo {author} {\bibfnamefont {A.}~\bibnamefont {Lane}}, \
  and\ \bibinfo {author} {\bibfnamefont {J.}~\bibnamefont {Martorell}},\ }\href
  {\doibase http://dx.doi.org/10.1016/0370-1573(79)90079-6} {\bibfield
  {journal} {\bibinfo  {journal} {Physics Reports}\ }\textbf {\bibinfo {volume}
  {51}},\ \bibinfo {pages} {267 } (\bibinfo {year} {1979})}\BibitemShut
  {NoStop}%
\bibitem [{\citenamefont {Capelli}\ \emph {et~al.}(2009)\citenamefont
  {Capelli}, \citenamefont {Col\`o},\ and\ \citenamefont {Li}}]{capelli2009}%
  \BibitemOpen
  \bibfield  {author} {\bibinfo {author} {\bibfnamefont {L.}~\bibnamefont
  {Capelli}}, \bibinfo {author} {\bibfnamefont {G.}~\bibnamefont {Col\`o}}, \
  and\ \bibinfo {author} {\bibfnamefont {J.}~\bibnamefont {Li}},\ }\href
  {\doibase 10.1103/PhysRevC.79.054329} {\bibfield  {journal} {\bibinfo
  {journal} {Phys. Rev. C}\ }\textbf {\bibinfo {volume} {79}},\ \bibinfo
  {pages} {054329} (\bibinfo {year} {2009})}\BibitemShut {NoStop}%
\bibitem [{\citenamefont {Hinohara}\ \emph {et~al.}(2015)\citenamefont
  {Hinohara}, \citenamefont {Kortelainen}, \citenamefont {Nazarewicz},\ and\
  \citenamefont {Olsen}}]{hinohara15}%
  \BibitemOpen
  \bibfield  {author} {\bibinfo {author} {\bibfnamefont {N.}~\bibnamefont
  {Hinohara}}, \bibinfo {author} {\bibfnamefont {M.}~\bibnamefont
  {Kortelainen}}, \bibinfo {author} {\bibfnamefont {W.}~\bibnamefont
  {Nazarewicz}}, \ and\ \bibinfo {author} {\bibfnamefont {E.}~\bibnamefont
  {Olsen}},\ }\href {\doibase 10.1103/PhysRevC.91.044323} {\bibfield  {journal}
  {\bibinfo  {journal} {Phys. Rev. C}\ }\textbf {\bibinfo {volume} {91}},\
  \bibinfo {pages} {044323} (\bibinfo {year} {2015})}\BibitemShut {NoStop}%
\bibitem [{\citenamefont {Myers}\ and\ \citenamefont
  {Swiatecki}(1974)}]{myers74}%
  \BibitemOpen
  \bibfield  {author} {\bibinfo {author} {\bibfnamefont {W.}~\bibnamefont
  {Myers}}\ and\ \bibinfo {author} {\bibfnamefont {W.}~\bibnamefont
  {Swiatecki}},\ }\href {\doibase 10.1016/0003-4916(74)90299-1} {\bibfield
  {journal} {\bibinfo  {journal} {Annals of Physics}\ }\textbf {\bibinfo
  {volume} {84}},\ \bibinfo {pages} {186 } (\bibinfo {year}
  {1974})}\BibitemShut {NoStop}%
\bibitem [{\citenamefont {Meyer}\ \emph {et~al.}(1982)\citenamefont {Meyer},
  \citenamefont {Quentin},\ and\ \citenamefont {Jennings}}]{meyer82}%
  \BibitemOpen
  \bibfield  {author} {\bibinfo {author} {\bibfnamefont {J.}~\bibnamefont
  {Meyer}}, \bibinfo {author} {\bibfnamefont {P.}~\bibnamefont {Quentin}}, \
  and\ \bibinfo {author} {\bibfnamefont {B.}~\bibnamefont {Jennings}},\ }\href
  {\doibase 10.1016/0375-9474(82)90172-5} {\bibfield  {journal} {\bibinfo
  {journal} {Nuclear Physics A}\ }\textbf {\bibinfo {volume} {385}},\ \bibinfo
  {pages} {269 } (\bibinfo {year} {1982})}\BibitemShut {NoStop}%
\bibitem [{\citenamefont {Warda}\ \emph {et~al.}(2009)\citenamefont {Warda},
  \citenamefont {Vi\~nas}, \citenamefont {Roca-Maza},\ and\ \citenamefont
  {Centelles}}]{warda09}%
  \BibitemOpen
  \bibfield  {author} {\bibinfo {author} {\bibfnamefont {M.}~\bibnamefont
  {Warda}}, \bibinfo {author} {\bibfnamefont {X.}~\bibnamefont {Vi\~nas}},
  \bibinfo {author} {\bibfnamefont {X.}~\bibnamefont {Roca-Maza}}, \ and\
  \bibinfo {author} {\bibfnamefont {M.}~\bibnamefont {Centelles}},\ }\href
  {\doibase 10.1103/PhysRevC.80.024316} {\bibfield  {journal} {\bibinfo
  {journal} {Phys. Rev. C}\ }\textbf {\bibinfo {volume} {80}},\ \bibinfo
  {pages} {024316} (\bibinfo {year} {2009})}\BibitemShut {NoStop}%
\bibitem [{\citenamefont {Satu\l{}a}\ \emph {et~al.}(2006)\citenamefont
  {Satu\l{}a}, \citenamefont {Wyss},\ and\ \citenamefont
  {Rafalski}}]{satula06}%
  \BibitemOpen
  \bibfield  {author} {\bibinfo {author} {\bibfnamefont {W.}~\bibnamefont
  {Satu\l{}a}}, \bibinfo {author} {\bibfnamefont {R.~A.}\ \bibnamefont {Wyss}},
  \ and\ \bibinfo {author} {\bibfnamefont {M.}~\bibnamefont {Rafalski}},\
  }\href {\doibase 10.1103/PhysRevC.74.011301} {\bibfield  {journal} {\bibinfo
  {journal} {Phys. Rev. C}\ }\textbf {\bibinfo {volume} {74}},\ \bibinfo
  {pages} {011301} (\bibinfo {year} {2006})}\BibitemShut {NoStop}%
\bibitem [{\citenamefont {Myers}\ and\ \citenamefont
  {Swiatecki}(1980)}]{myers1980}%
  \BibitemOpen
  \bibfield  {author} {\bibinfo {author} {\bibfnamefont {W.}~\bibnamefont
  {Myers}}\ and\ \bibinfo {author} {\bibfnamefont {W.}~\bibnamefont
  {Swiatecki}},\ }\href {\doibase
  http://dx.doi.org/10.1016/0375-9474(80)90623-5} {\bibfield  {journal}
  {\bibinfo  {journal} {Nuclear Physics A}\ }\textbf {\bibinfo {volume}
  {336}},\ \bibinfo {pages} {267 } (\bibinfo {year} {1980})}\BibitemShut
  {NoStop}%
\bibitem [{\citenamefont {Centelles}\ \emph {et~al.}(2009)\citenamefont
  {Centelles}, \citenamefont {Roca-Maza}, \citenamefont {Vi\~nas},\ and\
  \citenamefont {Warda}}]{centelles09}%
  \BibitemOpen
  \bibfield  {author} {\bibinfo {author} {\bibfnamefont {M.}~\bibnamefont
  {Centelles}}, \bibinfo {author} {\bibfnamefont {X.}~\bibnamefont
  {Roca-Maza}}, \bibinfo {author} {\bibfnamefont {X.}~\bibnamefont {Vi\~nas}},
  \ and\ \bibinfo {author} {\bibfnamefont {M.}~\bibnamefont {Warda}},\ }\href
  {\doibase 10.1103/PhysRevLett.102.122502} {\bibfield  {journal} {\bibinfo
  {journal} {Phys. Rev. Lett.}\ }\textbf {\bibinfo {volume} {102}},\ \bibinfo
  {pages} {122502} (\bibinfo {year} {2009})}\BibitemShut {NoStop}%
\bibitem [{\citenamefont {Centelles}\ \emph {et~al.}(2010)\citenamefont
  {Centelles}, \citenamefont {Roca-Maza}, \citenamefont {Vi\~nas},\ and\
  \citenamefont {Warda}}]{centelles10}%
  \BibitemOpen
  \bibfield  {author} {\bibinfo {author} {\bibfnamefont {M.}~\bibnamefont
  {Centelles}}, \bibinfo {author} {\bibfnamefont {X.}~\bibnamefont
  {Roca-Maza}}, \bibinfo {author} {\bibfnamefont {X.}~\bibnamefont {Vi\~nas}},
  \ and\ \bibinfo {author} {\bibfnamefont {M.}~\bibnamefont {Warda}},\ }\href
  {\doibase 10.1103/PhysRevC.82.054314} {\bibfield  {journal} {\bibinfo
  {journal} {Phys. Rev. C}\ }\textbf {\bibinfo {volume} {82}},\ \bibinfo
  {pages} {054314} (\bibinfo {year} {2010})}\BibitemShut {NoStop}%
\bibitem [{\citenamefont {Alex~Brown}(2000)}]{brown00}%
  \BibitemOpen
  \bibfield  {author} {\bibinfo {author} {\bibfnamefont {B.}~\bibnamefont
  {Alex~Brown}},\ }\href {\doibase 10.1103/PhysRevLett.85.5296} {\bibfield
  {journal} {\bibinfo  {journal} {Phys. Rev. Lett.}\ }\textbf {\bibinfo
  {volume} {85}},\ \bibinfo {pages} {5296} (\bibinfo {year}
  {2000})}\BibitemShut {NoStop}%
\bibitem [{\citenamefont {Furnstahl}(2002)}]{furnstahl02}%
  \BibitemOpen
  \bibfield  {author} {\bibinfo {author} {\bibfnamefont {R.}~\bibnamefont
  {Furnstahl}},\ }\href {\doibase 10.1016/S0375-9474(02)00867-9} {\bibfield
  {journal} {\bibinfo  {journal} {Nuclear Physics A}\ }\textbf {\bibinfo
  {volume} {706}},\ \bibinfo {pages} {85 } (\bibinfo {year}
  {2002})}\BibitemShut {NoStop}%
\bibitem [{\citenamefont {Todd}\ and\ \citenamefont
  {Piekarewicz}(2003)}]{Todd:2003xs}%
  \BibitemOpen
  \bibfield  {author} {\bibinfo {author} {\bibfnamefont {B.~G.}\ \bibnamefont
  {Todd}}\ and\ \bibinfo {author} {\bibfnamefont {J.}~\bibnamefont
  {Piekarewicz}},\ }\href@noop {} {\bibfield  {journal} {\bibinfo  {journal}
  {Phys. Rev.}\ }\textbf {\bibinfo {volume} {C67}},\ \bibinfo {pages} {044317}
  (\bibinfo {year} {2003})}\BibitemShut {NoStop}%
\bibitem [{\citenamefont {Sil}\ \emph {et~al.}(2005)\citenamefont {Sil},
  \citenamefont {Centelles}, \citenamefont {Vi\~nas},\ and\ \citenamefont
  {Piekarewicz}}]{sil05}%
  \BibitemOpen
  \bibfield  {author} {\bibinfo {author} {\bibfnamefont {T.}~\bibnamefont
  {Sil}}, \bibinfo {author} {\bibfnamefont {M.}~\bibnamefont {Centelles}},
  \bibinfo {author} {\bibfnamefont {X.}~\bibnamefont {Vi\~nas}}, \ and\
  \bibinfo {author} {\bibfnamefont {J.}~\bibnamefont {Piekarewicz}},\ }\href
  {\doibase 10.1103/PhysRevC.71.045502} {\bibfield  {journal} {\bibinfo
  {journal} {Phys. Rev. C}\ }\textbf {\bibinfo {volume} {71}},\ \bibinfo
  {pages} {045502} (\bibinfo {year} {2005})}\BibitemShut {NoStop}%
\bibitem [{dom()}]{dominic}%
  \BibitemOpen
  \href@noop {} {}\bibinfo {note} {Private communication with D. M.
  Rossi.}\BibitemShut {Stop}%
\bibitem [{\citenamefont {Niu}\ \emph {et~al.}(2014)\citenamefont {Niu},
  \citenamefont {Col\`o},\ and\ \citenamefont {Vigezzi}}]{niu15}%
  \BibitemOpen
  \bibfield  {author} {\bibinfo {author} {\bibfnamefont {Y.~F.}\ \bibnamefont
  {Niu}}, \bibinfo {author} {\bibfnamefont {G.}~\bibnamefont {Col\`o}}, \ and\
  \bibinfo {author} {\bibfnamefont {E.}~\bibnamefont {Vigezzi}},\ }\href
  {\doibase 10.1103/PhysRevC.90.054328} {\bibfield  {journal} {\bibinfo
  {journal} {Phys. Rev. C}\ }\textbf {\bibinfo {volume} {90}},\ \bibinfo
  {pages} {054328} (\bibinfo {year} {2014})}\BibitemShut {NoStop}%
\bibitem [{\citenamefont {Litvinova}(2015)}]{litvinova15}%
  \BibitemOpen
  \bibfield  {author} {\bibinfo {author} {\bibfnamefont {E.}~\bibnamefont
  {Litvinova}},\ }\href {\doibase 10.1103/PhysRevC.91.034332} {\bibfield
  {journal} {\bibinfo  {journal} {Phys. Rev. C}\ }\textbf {\bibinfo {volume}
  {91}},\ \bibinfo {pages} {034332} (\bibinfo {year} {2015})}\BibitemShut
  {NoStop}%
\bibitem [{\citenamefont {Lyutorovich}\ \emph {et~al.}(2015)\citenamefont
  {Lyutorovich}, \citenamefont {Tselyaev}, \citenamefont {Speth}, \citenamefont
  {Krewald}, \citenamefont {Grümmer},\ and\ \citenamefont
  {Reinhard}}]{lyutorovich15}%
  \BibitemOpen
  \bibfield  {author} {\bibinfo {author} {\bibfnamefont {N.}~\bibnamefont
  {Lyutorovich}}, \bibinfo {author} {\bibfnamefont {V.}~\bibnamefont
  {Tselyaev}}, \bibinfo {author} {\bibfnamefont {J.}~\bibnamefont {Speth}},
  \bibinfo {author} {\bibfnamefont {S.}~\bibnamefont {Krewald}}, \bibinfo
  {author} {\bibfnamefont {F.}~\bibnamefont {Grümmer}}, \ and\ \bibinfo
  {author} {\bibfnamefont {P.-G.}\ \bibnamefont {Reinhard}},\ }\href {\doibase
  http://dx.doi.org/10.1016/j.physletb.2015.08.003} {\bibfield  {journal}
  {\bibinfo  {journal} {Physics Letters B}\ }\textbf {\bibinfo {volume}
  {749}},\ \bibinfo {pages} {292 } (\bibinfo {year} {2015})}\BibitemShut
  {NoStop}%
\bibitem [{\citenamefont {Roca-Maza}\ \emph
  {et~al.}(2012{\natexlab{b}})\citenamefont {Roca-Maza}, \citenamefont
  {Col\`o},\ and\ \citenamefont {Sagawa}}]{roca-maza12b}%
  \BibitemOpen
  \bibfield  {author} {\bibinfo {author} {\bibfnamefont {X.}~\bibnamefont
  {Roca-Maza}}, \bibinfo {author} {\bibfnamefont {G.}~\bibnamefont {Col\`o}}, \
  and\ \bibinfo {author} {\bibfnamefont {H.}~\bibnamefont {Sagawa}},\ }\href
  {\doibase 10.1103/PhysRevC.86.031306} {\bibfield  {journal} {\bibinfo
  {journal} {Phys. Rev. C}\ }\textbf {\bibinfo {volume} {86}},\ \bibinfo
  {pages} {031306} (\bibinfo {year} {2012}{\natexlab{b}})}\BibitemShut
  {NoStop}%
\bibitem [{\citenamefont {Roca-Maza~{\it et al.}}(2013)}]{roca-maza13}%
  \BibitemOpen
  \bibfield  {author} {\bibinfo {author} {\bibfnamefont {X.}~\bibnamefont
  {Roca-Maza~{\it et al.}}},\ }\href {\doibase 10.1103/PhysRevC.87.034301}
  {\bibfield  {journal} {\bibinfo  {journal} {Phys. Rev. C}\ }\textbf {\bibinfo
  {volume} {87}},\ \bibinfo {pages} {034301} (\bibinfo {year}
  {2013})}\BibitemShut {NoStop}%
\bibitem [{\citenamefont {Draper}\ and\ \citenamefont
  {Smith}(1998)}]{draper81}%
  \BibitemOpen
  \bibfield  {author} {\bibinfo {author} {\bibfnamefont {N.}~\bibnamefont
  {Draper}}\ and\ \bibinfo {author} {\bibfnamefont {H.}~\bibnamefont {Smith}},\
  }\href@noop {} {\emph {\bibinfo {title} {Applied Regression Analysis}}}\
  (\bibinfo  {publisher} {Wiley, New York},\ \bibinfo {year}
  {1998})\BibitemShut {NoStop}%
\bibitem [{\citenamefont {Bender}\ \emph {et~al.}(2003)\citenamefont {Bender},
  \citenamefont {Heenen},\ and\ \citenamefont {Reinhard}}]{skyrme1}%
  \BibitemOpen
  \bibfield  {author} {\bibinfo {author} {\bibfnamefont {M.}~\bibnamefont
  {Bender}}, \bibinfo {author} {\bibfnamefont {P.-H.}\ \bibnamefont {Heenen}},
  \ and\ \bibinfo {author} {\bibfnamefont {P.-G.}\ \bibnamefont {Reinhard}},\
  }\href {\doibase 10.1103/RevModPhys.75.121} {\bibfield  {journal} {\bibinfo
  {journal} {Rev. Mod. Phys.}\ }\textbf {\bibinfo {volume} {75}},\ \bibinfo
  {pages} {121} (\bibinfo {year} {2003})}\BibitemShut {NoStop}%
\bibitem [{\citenamefont {Reinhard}\ \emph {et~al.}(2006)\citenamefont
  {Reinhard}, \citenamefont {Bender}, \citenamefont {Nazarewicz},\ and\
  \citenamefont {Vertse}}]{skyrme2}%
  \BibitemOpen
  \bibfield  {author} {\bibinfo {author} {\bibfnamefont {P.-G.}\ \bibnamefont
  {Reinhard}}, \bibinfo {author} {\bibfnamefont {M.}~\bibnamefont {Bender}},
  \bibinfo {author} {\bibfnamefont {W.}~\bibnamefont {Nazarewicz}}, \ and\
  \bibinfo {author} {\bibfnamefont {T.}~\bibnamefont {Vertse}},\ }\href
  {\doibase 10.1103/PhysRevC.73.014309} {\bibfield  {journal} {\bibinfo
  {journal} {Phys. Rev. C}\ }\textbf {\bibinfo {volume} {73}},\ \bibinfo
  {pages} {014309} (\bibinfo {year} {2006})}\BibitemShut {NoStop}%
\bibitem [{\citenamefont {Cao}\ \emph {et~al.}(2006)\citenamefont {Cao},
  \citenamefont {Lombardo}, \citenamefont {Shen},\ and\ \citenamefont
  {Giai}}]{lns}%
  \BibitemOpen
  \bibfield  {author} {\bibinfo {author} {\bibfnamefont {L.~G.}\ \bibnamefont
  {Cao}}, \bibinfo {author} {\bibfnamefont {U.}~\bibnamefont {Lombardo}},
  \bibinfo {author} {\bibfnamefont {C.~W.}\ \bibnamefont {Shen}}, \ and\
  \bibinfo {author} {\bibfnamefont {N.~V.}\ \bibnamefont {Giai}},\ }\href
  {\doibase 10.1103/PhysRevC.73.014313} {\bibfield  {journal} {\bibinfo
  {journal} {Phys. Rev. C}\ }\textbf {\bibinfo {volume} {73}},\ \bibinfo
  {pages} {014313} (\bibinfo {year} {2006})}\BibitemShut {NoStop}%
\bibitem [{\citenamefont {Agrawal}\ \emph {et~al.}(2003)\citenamefont
  {Agrawal}, \citenamefont {Shlomo},\ and\ \citenamefont {Kim~Au}}]{sk255}%
  \BibitemOpen
  \bibfield  {author} {\bibinfo {author} {\bibfnamefont {B.~K.}\ \bibnamefont
  {Agrawal}}, \bibinfo {author} {\bibfnamefont {S.}~\bibnamefont {Shlomo}}, \
  and\ \bibinfo {author} {\bibfnamefont {V.}~\bibnamefont {Kim~Au}},\ }\href
  {\doibase 10.1103/PhysRevC.68.031304} {\bibfield  {journal} {\bibinfo
  {journal} {Phys. Rev. C}\ }\textbf {\bibinfo {volume} {68}},\ \bibinfo
  {pages} {031304} (\bibinfo {year} {2003})}\BibitemShut {NoStop}%
\bibitem [{\citenamefont {Agrawal}\ \emph {et~al.}(2005)\citenamefont
  {Agrawal}, \citenamefont {Shlomo},\ and\ \citenamefont {Au}}]{kde0}%
  \BibitemOpen
  \bibfield  {author} {\bibinfo {author} {\bibfnamefont {B.~K.}\ \bibnamefont
  {Agrawal}}, \bibinfo {author} {\bibfnamefont {S.}~\bibnamefont {Shlomo}}, \
  and\ \bibinfo {author} {\bibfnamefont {V.~K.}\ \bibnamefont {Au}},\ }\href
  {\doibase 10.1103/PhysRevC.72.014310} {\bibfield  {journal} {\bibinfo
  {journal} {Phys. Rev. C}\ }\textbf {\bibinfo {volume} {72}},\ \bibinfo
  {pages} {014310} (\bibinfo {year} {2005})}\BibitemShut {NoStop}%
\bibitem [{kde()}]{kde0j}%
  \BibitemOpen
  \href@noop {} {}\bibinfo {note} {KDE0-J family was recently generated by one
  of the authors. The details have not been published yet.}\BibitemShut {Stop}%
\bibitem [{\citenamefont {Lalazissis}\ \emph {et~al.}(1997)\citenamefont
  {Lalazissis}, \citenamefont {K\"onig},\ and\ \citenamefont
  {Ring}}]{Lalazissis:1996rd}%
  \BibitemOpen
  \bibfield  {author} {\bibinfo {author} {\bibfnamefont {G.~A.}\ \bibnamefont
  {Lalazissis}}, \bibinfo {author} {\bibfnamefont {J.}~\bibnamefont {K\"onig}},
  \ and\ \bibinfo {author} {\bibfnamefont {P.}~\bibnamefont {Ring}},\ }\href
  {\doibase 10.1103/PhysRevC.55.540} {\bibfield  {journal} {\bibinfo  {journal}
  {Phys. Rev. C}\ }\textbf {\bibinfo {volume} {55}},\ \bibinfo {pages} {540}
  (\bibinfo {year} {1997})}\BibitemShut {NoStop}%
\bibitem [{\citenamefont {Agrawal}(2010)}]{nl3fsu1}%
  \BibitemOpen
  \bibfield  {author} {\bibinfo {author} {\bibfnamefont {B.~K.}\ \bibnamefont
  {Agrawal}},\ }\href {\doibase 10.1103/PhysRevC.81.034323} {\bibfield
  {journal} {\bibinfo  {journal} {Phys. Rev. C}\ }\textbf {\bibinfo {volume}
  {81}},\ \bibinfo {pages} {034323} (\bibinfo {year} {2010})}\BibitemShut
  {NoStop}%
\bibitem [{\citenamefont {Fattoyev}\ and\ \citenamefont
  {Piekarewicz}(2013)}]{fattoyev13}%
  \BibitemOpen
  \bibfield  {author} {\bibinfo {author} {\bibfnamefont {F.~J.}\ \bibnamefont
  {Fattoyev}}\ and\ \bibinfo {author} {\bibfnamefont {J.}~\bibnamefont
  {Piekarewicz}},\ }\href {\doibase 10.1103/PhysRevLett.111.162501} {\bibfield
  {journal} {\bibinfo  {journal} {Phys. Rev. Lett.}\ }\textbf {\bibinfo
  {volume} {111}},\ \bibinfo {pages} {162501} (\bibinfo {year}
  {2013})}\BibitemShut {NoStop}%
\bibitem [{\citenamefont {Ebata}\ \emph {et~al.}(2014)\citenamefont {Ebata},
  \citenamefont {Nakatsukasa},\ and\ \citenamefont {Inakura}}]{Ebata2014}%
  \BibitemOpen
  \bibfield  {author} {\bibinfo {author} {\bibfnamefont {S.}~\bibnamefont
  {Ebata}}, \bibinfo {author} {\bibfnamefont {T.}~\bibnamefont {Nakatsukasa}},
  \ and\ \bibinfo {author} {\bibfnamefont {T.}~\bibnamefont {Inakura}},\ }\href
  {\doibase 10.1103/PhysRevC.90.024303} {\bibfield  {journal} {\bibinfo
  {journal} {Phys. Rev. C}\ }\textbf {\bibinfo {volume} {90}},\ \bibinfo
  {pages} {024303} (\bibinfo {year} {2014})}\BibitemShut {NoStop}%
\bibitem [{\citenamefont {Tsang~{\it et al.}}(2012)}]{tsang12}%
  \BibitemOpen
  \bibfield  {author} {\bibinfo {author} {\bibfnamefont {M.~B.}\ \bibnamefont
  {Tsang~{\it et al.}}},\ }\href {\doibase 10.1103/PhysRevC.86.015803}
  {\bibfield  {journal} {\bibinfo  {journal} {Phys. Rev. C}\ }\textbf {\bibinfo
  {volume} {86}},\ \bibinfo {pages} {015803} (\bibinfo {year}
  {2012})}\BibitemShut {NoStop}%
\bibitem [{\citenamefont {Lattimer}\ and\ \citenamefont
  {Lim}(2013)}]{lattimer2013}%
  \BibitemOpen
  \bibfield  {author} {\bibinfo {author} {\bibfnamefont {J.~M.}\ \bibnamefont
  {Lattimer}}\ and\ \bibinfo {author} {\bibfnamefont {Y.}~\bibnamefont {Lim}},\
  }\href {http://stacks.iop.org/0004-637X/771/i=1/a=51} {\bibfield  {journal}
  {\bibinfo  {journal} {The Astrophysical Journal}\ }\textbf {\bibinfo {volume}
  {771}},\ \bibinfo {pages} {51} (\bibinfo {year} {2013})}\BibitemShut
  {NoStop}%
\bibitem [{\citenamefont {Blaizot}\ \emph {et~al.}(1995)\citenamefont
  {Blaizot}, \citenamefont {Berger}, \citenamefont {Decharge},\ and\
  \citenamefont {Girod}}]{Blaizot:1995zz}%
  \BibitemOpen
  \bibfield  {author} {\bibinfo {author} {\bibfnamefont {J.}~\bibnamefont
  {Blaizot}}, \bibinfo {author} {\bibfnamefont {J.}~\bibnamefont {Berger}},
  \bibinfo {author} {\bibfnamefont {J.}~\bibnamefont {Decharge}}, \ and\
  \bibinfo {author} {\bibfnamefont {M.}~\bibnamefont {Girod}},\ }\href
  {\doibase 10.1016/0375-9474(95)00294-B} {\bibfield  {journal} {\bibinfo
  {journal} {Nucl.Phys.}\ }\textbf {\bibinfo {volume} {A591}},\ \bibinfo
  {pages} {435} (\bibinfo {year} {1995})}\BibitemShut {NoStop}%
%%CITATION = NUPHA,A591,435;%%
\bibitem [{\citenamefont {Dutra}\ \emph {et~al.}(2012)\citenamefont {Dutra},
  \citenamefont {Louren\ifmmode~\mbox{\c{c}}\else \c{c}\fi{}o}, \citenamefont
  {S\'a~Martins}, \citenamefont {Delfino}, \citenamefont {Stone},\ and\
  \citenamefont {Stevenson}}]{dutra12}%
  \BibitemOpen
  \bibfield  {author} {\bibinfo {author} {\bibfnamefont {M.}~\bibnamefont
  {Dutra}}, \bibinfo {author} {\bibfnamefont {O.}~\bibnamefont
  {Louren\ifmmode~\mbox{\c{c}}\else \c{c}\fi{}o}}, \bibinfo {author}
  {\bibfnamefont {J.~S.}\ \bibnamefont {S\'a~Martins}}, \bibinfo {author}
  {\bibfnamefont {A.}~\bibnamefont {Delfino}}, \bibinfo {author} {\bibfnamefont
  {J.~R.}\ \bibnamefont {Stone}}, \ and\ \bibinfo {author} {\bibfnamefont
  {P.~D.}\ \bibnamefont {Stevenson}},\ }\href {\doibase
  10.1103/PhysRevC.85.035201} {\bibfield  {journal} {\bibinfo  {journal} {Phys.
  Rev. C}\ }\textbf {\bibinfo {volume} {85}},\ \bibinfo {pages} {035201}
  (\bibinfo {year} {2012})}\BibitemShut {NoStop}%
\bibitem [{\citenamefont {Vi\~nas}\ \emph {et~al.}(2014)\citenamefont
  {Vi\~nas}, \citenamefont {Centelles}, \citenamefont {Roca-Maza},\ and\
  \citenamefont {Warda}}]{vinas14}%
  \BibitemOpen
  \bibfield  {author} {\bibinfo {author} {\bibfnamefont {X.}~\bibnamefont
  {Vi\~nas}}, \bibinfo {author} {\bibfnamefont {M.}~\bibnamefont {Centelles}},
  \bibinfo {author} {\bibfnamefont {X.}~\bibnamefont {Roca-Maza}}, \ and\
  \bibinfo {author} {\bibfnamefont {M.}~\bibnamefont {Warda}},\ }\href
  {\doibase 10.1140/epja/i2014-14027-8} {\bibfield  {journal} {\bibinfo
  {journal} {The European Physical Journal A}\ }\textbf {\bibinfo {volume}
  {50}},\ \bibinfo {eid} {27} (\bibinfo {year} {2014}),\
  10.1140/epja/i2014-14027-8}\BibitemShut {NoStop}%
\bibitem [{\citenamefont {Li}\ and\ \citenamefont {Han}(2013)}]{bao13}%
  \BibitemOpen
  \bibfield  {author} {\bibinfo {author} {\bibfnamefont {B.-A.}\ \bibnamefont
  {Li}}\ and\ \bibinfo {author} {\bibfnamefont {X.}~\bibnamefont {Han}},\
  }\href {\doibase http://dx.doi.org/10.1016/j.physletb.2013.10.006} {\bibfield
   {journal} {\bibinfo  {journal} {Physics Letters B}\ }\textbf {\bibinfo
  {volume} {727}},\ \bibinfo {pages} {276 } (\bibinfo {year}
  {2013})}\BibitemShut {NoStop}%
\bibitem [{\citenamefont {Col\`o}\ \emph {et~al.}(2014)\citenamefont {Col\`o},
  \citenamefont {Garg},\ and\ \citenamefont {Sagawa}}]{colo15}%
  \BibitemOpen
  \bibfield  {author} {\bibinfo {author} {\bibfnamefont {G.}~\bibnamefont
  {Col\`o}}, \bibinfo {author} {\bibfnamefont {U.}~\bibnamefont {Garg}}, \ and\
  \bibinfo {author} {\bibfnamefont {H.}~\bibnamefont {Sagawa}},\ }\href
  {\doibase 10.1140/epja/i2014-14026-9} {\bibfield  {journal} {\bibinfo
  {journal} {The European Physical Journal A}\ }\textbf {\bibinfo {volume}
  {50}},\ \bibinfo {pages} {26} (\bibinfo {year} {2014})}\BibitemShut {NoStop}%
\bibitem [{\citenamefont {Roca-Maza}\ \emph {et~al.}(2011)\citenamefont
  {Roca-Maza}, \citenamefont {Centelles}, \citenamefont {Vi\~nas},\ and\
  \citenamefont {Warda}}]{roca-maza11}%
  \BibitemOpen
  \bibfield  {author} {\bibinfo {author} {\bibfnamefont {X.}~\bibnamefont
  {Roca-Maza}}, \bibinfo {author} {\bibfnamefont {M.}~\bibnamefont
  {Centelles}}, \bibinfo {author} {\bibfnamefont {X.}~\bibnamefont {Vi\~nas}},
  \ and\ \bibinfo {author} {\bibfnamefont {M.}~\bibnamefont {Warda}},\ }\href
  {\doibase 10.1103/PhysRevLett.106.252501} {\bibfield  {journal} {\bibinfo
  {journal} {Phys. Rev. Lett.}\ }\textbf {\bibinfo {volume} {106}},\ \bibinfo
  {pages} {252501} (\bibinfo {year} {2011})}\BibitemShut {NoStop}%
\bibitem [{\citenamefont {Horowitz}\ \emph {et~al.}(2014)\citenamefont
  {Horowitz}, \citenamefont {Brown}, \citenamefont {Kim}, \citenamefont
  {Lynch}, \citenamefont {Michaels} \emph {et~al.}}]{Horowitz:2014bja}%
  \BibitemOpen
  \bibfield  {author} {\bibinfo {author} {\bibfnamefont {C.~J.}\ \bibnamefont
  {Horowitz}}, \bibinfo {author} {\bibfnamefont {E.~F.}\ \bibnamefont {Brown}},
  \bibinfo {author} {\bibfnamefont {Y.}~\bibnamefont {Kim}}, \bibinfo {author}
  {\bibfnamefont {W.~G.}\ \bibnamefont {Lynch}}, \bibinfo {author}
  {\bibfnamefont {R.}~\bibnamefont {Michaels}},  \emph {et~al.},\ }\href@noop
  {} {\bibfield  {journal} {\bibinfo  {journal} {J. Phys.}\ }\textbf {\bibinfo
  {volume} {G41}},\ \bibinfo {pages} {093001} (\bibinfo {year}
  {2014})}\BibitemShut {NoStop}%
\end{thebibliography}%
\end{document}